\documentstyle[epsfig,MA767,twocolumn]{mn}

 \title[Super Eddington Winds]{\sc The Theory of Steady State Super-Eddington Winds and its
 Application to Novae}

 \author[N. J. Shaviv]{Nir J. Shaviv \\
        Canadian Institute for Theoretical Astrophysics, University of Toronto \\
        60 St. George St., Toronto, ON M5S 3H8, Canada}

\date{Accepted to MNRAS, CITA-2000-30}

 \pagerange{\pageref{firstpage}--\pageref{lastpage}}
 \pubyear{2001}

\def\nsk{\hskip -3pt}
\def\st{0}
\def\unit{{\mr{unit}}}
\def\sonic{{\mr{sonic}}}
\def\A{{\cal A}}
\def\B{{\cal B}}

\begin{document}
\label{firstpage}

\maketitle

\begin{abstract}

 We present a model for steady state winds of systems with
 super-Eddington luminosities. These radiatively driven winds are
 expected to be optically thick and clumpy as they arise from an
 instability driven porous atmosphere.  The model is then
 applied to derive the mass loss observed in bright classical
 novae. The main results are:

{\leftmargin 0pt
\parsep 0pt
\begin{enumerate}
 \item A general relation between the mass loss rate and the
 total luminosity in super-Eddington systems.

 \item A natural explanation to the long duration super-Eddington
 outflows that are clearly observed in at least two cases (Novae LMC
 1988 \#1 \& FH Serpentis).

 \item A quantitative agreement between the observed luminosity
 evolution which is used to predict both the mass loss and temperature
 evolution, and their observations.

 \item An agreement between the predicted average integrated mass loss of novae
 as a function of WD mass and its observations.

 \item A natural explanation for the `transition phase' of novae.

 \item Agreement with $\eta$ Carinae which was used to double check
 the theory. The prediction for the mass shed in the star's great
 eruption agrees with observations to within the measurement error.
\end{enumerate}
}
 
\end{abstract}

\begin{keywords}
  Radiative transfer --- hydrodynamics --- instabilities --- stars:
  atmospheres --- stars: individual LMC 1988 \#1, FH Ser,
  $\eta$ Carinae --- novae, cataclysmic variables
\end{keywords}

\section{Introduction}

 The Eddington luminosity is the maximum luminosity allowed for a {\em
 stationary} spherical, homogeneous, non-relativistic and fully
 ionised system. If one allows motion, then a {\em steady state}
 super-Eddington configuration generally does not exist unless the
 system is just marginally super-Eddington or it has a very high mass
 loss rate. Nevertheless, nature does find a way to construct steady
 state configurations in which super-Eddington luminosities exist with
 only a relatively small mass loss rate. This was perhaps best
 demonstrated with $\eta$ Carinae's giant eruption.

 $\eta$ Carinae was clearly super-Eddington during its 20 year long
 eruption (see for instance the review by \cite{Davidson97}). Yet, it
 was shown that the observed mass loss and velocity are inconsistent
 with a homogeneous solution for the wind (\cite{etacar}). Basically,
 the sonic point obtained from the observed conditions necessarily has
 to reside too high in the atmosphere, at an optical depth of only
 $\sim 1$ to $\sim 300$, while the critical point in a homogeneous
 atmosphere necessarily has to reside at significantly deeper optical
 depths. The inconsistency arises because the sonic point and the
 critical point have to coincide in a steady state solution, while
 nowhere within the plausible space of parameters such a solution
 exists.

 A solution was proposed in which the atmosphere of $\eta$ Carinae is
 inhomogeneous, or porous (\cite{etacar}). The inhomogeneity is a
 natural result of the instabilities of atmospheres that are close to
 the Eddington luminosity (\cite{instabilities}). The
 inhomogeneities, or `porosity', reduce the effective opacity and
 increase the effective Eddington luminosity (\cite{porous}).

 In this paper, we are interested in understanding the wind generated
 in cases in which the luminosity is super-Eddington.  To do so, it is
 advantageous to find a class of objects for which better data than
 for $\eta$ Car exists. One such class of objects is novae.
 
 Novae are a very good population to analyze in order to understand
 the behaviour of steady state super-Eddington winds. The main reasons
 are:
\begin{enumerate}
 \item The mass and luminosity are better known than for many other
 objects. For example, although $\eta$ Carinae was clearly super
 Eddington, it is not clear by how much it was so: Its mass can be
 anywhere between $60 M_{\sun}$ and $100 M_{\sun}$ and its luminosity
 during the eruption is even less accurately known. The ejected mass
 could have been between $1 M_{\sun}$ and $3 M_{\sun}$. This is not
 accurate enough for our purposes here.

 \item The opacity where the sonic point is expected to be located is
 governed by Thomson scattering. In AGB and post-AGB stars that
 generate strong radiatively driven winds, the opacity is a very
 sensitive function of the local parameters of the gas at the sonic
 point. Thus, even though their observed mass and luminosity can be
 fairly accurately deduced from the observations, the {\em modified}
 Eddington parameter which should take into account the opacity of the
 dust, for example, is not known reasonably well.

 \item Since the luminosity during the super-Eddington episode of
 novae eruptions can be significantly above the Eddington limit, the
 inaccuracy of $\Gamma=L/\L_{\mr{Edd}}$ is less critical to the exact
 value of the mass loss.  In objects that shine very close to the
 Eddington limit for a long duration, with a relatively small mass
 loss rate, the theory cannot give precise predictions.  Thus, if for
 example the winds of the most luminous WR stars arises because the
 objects are marginally super-Eddington, it would be hard to compare
 their observations to the theory presented here because the accuracy
 in the determination of $\Gamma-1$ will be rather poor.
 
 \item Novae generally exhibit a `bolometric plateau' in which the
 bolometric luminosity decreases slowly over a relatively long
 duration, significantly longer than the dynamical time scale.
 Therefore, if this luminosity is super-Eddington, then clearly a
 steady state model for the super-Eddington flow should be
 constructed.  This is clearly the case with two specific novae:
 \NLMC~and \NFH. We exclude from the discussion here the very fast
 novae for which this property is least pronounced.
\end{enumerate}

 The fact that the `bolometric plateau' is sometimes observed to be
 super-Eddington does not presently have any good theoretical
 explanation. The steady state burning of the post maximum of novae is
 often predicted to be given or at least approximated by the core-mass
 luminosity relation (\cite{Paczynski70}). The classical core-mass
 luminosity relation increases monotonically with the mass of the WD
 and saturates at the Eddington limit.  It does not yield
 super-Eddington luminosities. The observations of super-Eddington
 luminosities over durations much longer than the dynamic time scale,
 lead us to the hypothesis that bright novae must have a porous
 atmosphere. A porous atmosphere with a reduced effective opacity will
 naturally give a core-mass luminosity relation that increases
 monotonically with mass beyond the classical Eddington limit,
 providing the arena for steady state super-Eddington winds.

 It is these winds which are the subject of this paper. In section
 \ref{sec:wind} we present the `wind theory' for super-Eddington
 atmospheres.  In section \ref{sec:novae}, we apply the wind theory to
 two specific novae that were clearly super-Eddington over a long
 duration, to the nova population in general and to $\eta$ Carinae and
 compare the results with observations.

\section{The Fundamental Structure of Super-Eddington Winds (SEWs)}
\label{sec:wind}
\subsection{Some general considerations}
 What have we learned from $\eta$ Carinae?  $\eta$ Car has shown us
 that an object can be super-Eddington for a duration much longer than
 the dynamical time scale while driving a wind which is significantly
 thinner than one should expect in a homogeneous wind solution.  When
 one tries to construct a steady state wind solution, one has to place
 the sonic point at the critical point---where the net forces on a
 mass element vanish.  If a system is in steady state and
 super-Eddington, then the critical point has to reside where
 alternative means of transporting the energy flux, namely by
 convection or advection with the flow, become inefficient.  This
 point however, happens relatively deep in the atmosphere ($\tau \gg
 1$), implying that the mass loss $\Md=4\pi R^2\rho v_s$ (where
 $v_{s}$ is the sound velocity), is very large.  In fact, the mass
 loss rate becomes of order $(L - \L_{\mr{Edd}})/v_s^2$ (e.g.,
 \cite{Owocki97}, \cite{etacar}).  However, if the radius of the
 system is fixed, then because a minimum energy flux of $\Md
 GM_\star/R = \Md v_{\mr{esc}}^2/2$ (with $v_{\mr{esc}}$ being the
 escape velocity) has to be supplied in order to pump the material out
 of the gravitational well, one obtains that unless $v_{s} \gtrsim
 v_{\mr{esc}}$, the radiation will not be able to provide the work
 needed (\cite{Owocki97}).  This implies that below a certain
 luminosity and radius of the system, there is no steady state
 solution.  Clearly, the system would try to expand its outer layers
 (that are driven outward but cannot reach infinity), thereby reducing
 the escape velocity, until a steady state can be reached.

 One would expect that as time progresses, the sonic point of the wind
 would move monotonically downwards, pushing more and more material
 upwards thereby expanding the atmosphere and accelerating more mass
 until all the available luminosity would be used-up to pump material
 out of the well.  If this expectation is realized, $\eta$ Car would
 have appeared as a faint object with high mass loss at low
 velocities.  In other words, observations of $\eta$ Car suggest that
 down to the optical depth of at most $\lesssim 300$, which is the
 deepest that the sonic point could be located (\cite{etacar}), the
 total mass is $\lesssim 0.02 M_{\sun}$, implying that this part of
 the envelope should have been in steady state for time scales longer
 than $\sim 3$ months.  This hypothetical steady state is inconsistent
 with $\eta $ Car trying to reach an equilibrium in which most of the
 radiation is used up to accelerate a very large amount of mass to low
 velocities, the star has been notably super-Eddington for a long
 duration without accelerating more and more mass at lower velocities.

 We will soon show that novae, at least during their `bolometric
 plateau' defy the Eddington limit. In some cases at least, a steady
 state super-Eddington configuration is reached in which the kinetic
 energy in the flow plus the rate in which gravitational energy is
 pumped is only a small or moderate fraction of the total radiative
 energy flux at the base of the wind. This too is inconsistent with a
 sonic point located deep inside the atmosphere.

 So, how did $\eta$ Car circumvent its bloating up?  It was proposed
 by Shaviv (\jcite{etacar}) that the solution to the problem is in
 having a porous atmosphere. In such an atmosphere, density
 perturbations naturally reduce the effective opacity
 $\kappa_{\mr{eff}}$ (\cite{porous}). Consequently, the radiative
 force is decreased and the effective Eddington limit is increased to:
\begin{equation}
\L_{\mr{eff}} = {\kappa_{0} \over \kappa_{\mr{eff}}} \L_{\mr{Edd}} >
\L_{\mr{Edd}},
\end{equation}
 where $\kappa_{0}$ is the microscopic opacity.  When
 $\kappa_{0}=\kappa_{\mr{Thomson}}$, the Eddington limit
 $\L_{\mr{Edd}}$ corresponds to the {\em classical} Eddington limit.
 In most other cases, where the microscopic opacity $\kappa_0$ is
 larger than the Thomson opacity, $\L_{\mr{Edd}}$ corresponds to a
 lower {\em modified} Eddington luminosity.  In the rest of this work,
 we shall not explicitly state whether the Eddington limit is
 `classical' or `modified'.  The distinction should be made according
 to the underlying microscopic opacity, which is close to that of
 Thomson scattering in the specific cases studied here.  We shall
 however make the important distinction between the `modified' and the
 `effective' Eddington luminosity, the latter being the effective
 Eddington luminosity in a non-homogeneous system.

 A mechanism which converts the homogeneous layers into inhomogeneous
 was suggested by Shaviv (\jcite{instabilities}). It is demonstrated
 that as the radiative flux through the atmospheres surpasses a
 critical Eddington parameter of $\Gamma_{\mr{crit}} \sim 0.5 - 0.85$
 (the exact numerical value depends on the boundary conditions), the
 atmosphere becomes unstable to at least two different instabilities,
 both of which operate on the dynamical time scale (namely, the sound
 crossing time of a scale height). Consequently, as the radiative flux
 approaches the Eddington limit and surpasses it, the radiation
 triggers the transition of the atmosphere from a homogeneous one to
 an inhomogeneous one.  The inhomogeneities increase the effective
 Eddington limit thereby functionally keeping the radiation flux at a
 sub-Eddington level; namely, even if $\Gamma>1$, the optically thick
 regions experience a $\Gamma_{\mr{eff}} < 1$. Other instabilities
 that operate in more complex environments could too be important and
 contribute to this transition. For example, $s$-mode instabilities
 (\cite{Glatzel93}) or the instability of dynamically detached outer
 layers (\cite{Stothers93}) operate under more complex opacity
 laws. The instability of `photon bubbles' appears when strong
 magnetic fields are present (\cite{Arons92}). In principle, since the
 adiabatic index approaches the critical value of 4/3 for instability
 as the Eddington limit is approached, many mechanisms which are
 otherwise unimportant do become important.

 The `porosity' reduces the effective opacity only as long as the
 perturbations are optically thick. Therefore, a wind will necessarily
 be generated from the regions in which the perturbations become
 optically thin, since from these regions upwards the effective
 opacity will be the normal `microscopic' one and the effective
 Eddington limit will return to be the classical one.

 {\em The $\Gamma_{\mr{crit}} < \Gamma < 1$ case}:---The details of
 the geometry (or inhomogeneities) of the regions depend on the
 instability at play, and for example can be in the form of `chimneys'
 or `photon bubbles'.  The lowered opacity is achieved by funneling
 the radiation through regions with a much lower than average density.
 These regions can be super-Eddington even when the mean $\Gamma$
 parameter is smaller than unity.  In such cases, mass loss should be
 driven in the super-Eddington `chimneys' of lower density.  These
 atmospheres are then expected to be very dynamic.  Once any
 accelerated mass element reaches the optically thin part of the
 atmosphere, it will start to experience an average force that is at a
 sub-Eddington value and the mass flow will stagnate, probably forming
 something which looks like `geysers'.  (It is very unlikely that the
 escape velocity will be attainable in the `chimneys' since shocks
 would probably limit the flow to velocities not much larger than the
 speed of sound).  A wind could then be generated in the optically
 thin part of the atmosphere through the standard line driving
 mechanism (\cite{CAK}, or for example \cite{Pauldrach86} and
 references therein), with the notable consideration that the base of
 the wind is clumpy.

 {\em The $\Gamma > 1$ case}:---When $\Gamma>1$ on the other hand, a
 continuum driven wind has to be generated.  The reason is clear.
 Since the perturbations have to be optically thick to affect (and
 reduce) the opacity, at a low enough optical depth one would expect
 to return back to a super-Eddington flow.  What is this optical
 depth?  If we climb up the thick yet porous atmosphere, since it is
 effectively sub-Eddington, the average density will decay
 exponentially with height.  At some point, the density will be low
 enough that a typical perturbation `element' becomes optically thin.
 Since the perturbations are expected to be of order a scale height in
 size (\cite{instabilities}), this depth would be where a scale height
 has an optical {\em width} of order unity\footnote{Note that if the
 atmosphere is static and therefore has an exponential density
 profile, this location would also correspond to the place at which
 the optical {\em depth} is of order unity.  Since a thick wind is
 expected to form, the physical depth where the optical {\em width} of
 a scale height is of order unity does not correspond to the physical
 location (the photosphere) where the optical {\em depth} is of order
 unity.}.  Beyond this point, the typical perturbation on scales of
 order the a scale height cannot reduce the opacity.  This is the
 place where the sonic point should be located for a steady wind to
 exist.

 More specifically, if the flux corresponds to an Eddington parameter
 $\Gamma$, then the optical depth at which perturbations cannot reduce
 the effective Eddington parameter to unity should scale with
 $\Gamma-1$.  The reason is that the decrease of the luminosity by the
 effective opacity should be proportional to the deviation of the
 actual luminosity from the Eddington one.  Namely, when close to the
 Eddington limit, a blob with the same geometry needs a smaller
 density fluctuation and with it a smaller change in the opacity, to
 reduce the effective opacity by the amount needed to become
 Eddington.  Thus, when closer to the Eddington limit, the sonic point
 can sit higher in the atmosphere.  We shall elaborate this point to
 show that this is indeed the case in \S\ref{sec:criticalpoint}

\subsection{The Structure of a Steady SEW}

\begin{figure}
\centerline{\epsfig{file=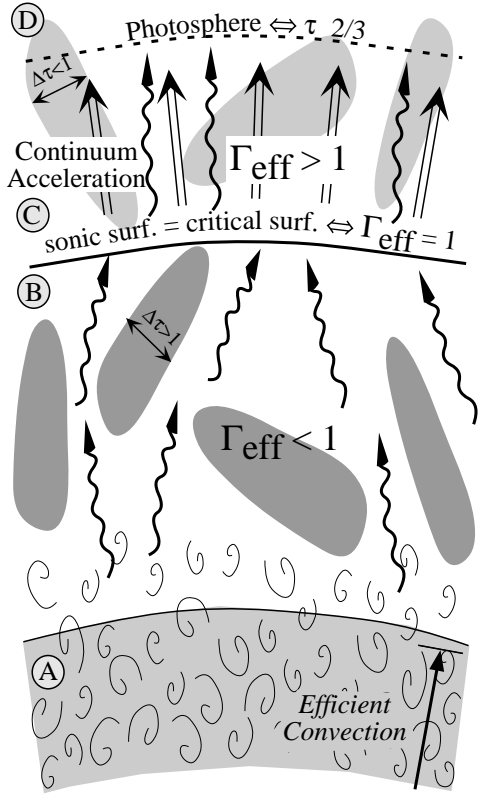,width=\figurewidthsmall}}
\caption{
     The proposed structure of a super-Eddington atmosphere (one with
     $L_{\mr{tot}}>\L_{\mr{Edd}}$) and the wind that it generates. The
     four regions, which are described in detail in the text, are:
     Region (A) A Convective envelope in which the radiation is
     sub-Eddington -- $L_{\mr{rad}} < \L_{\mr{Edd}} <
     L_{\mr{tot}}$. Region (B) A porous atmosphere in which the
     effective Eddington luminosity is larger than the classical
     Eddington luminosity: $\L_{\mr{Edd}} < L_{\mr{rad}} =
     L_{\mr{tot}} < \L_{\mr{eff}}$. Region (C) of an optically thick,
     continuum driven wind, where perturbations are optically thin and
     the effective Eddington limit tends to the classical
     value. Region (D) of a photosphere and above.}
\label{fig:structure}
\end{figure}


 The above considerations lead us to propose the following structure
 of a super-Eddington wind (hereafter SEW). Consider
 Fig.~\ref{fig:structure} for the proposed structure of a
 super-Eddington atmosphere (one with $L_{\mr{tot}}>\L_{\mr{Edd}}$)
 and the SEW that it generates.  Four main regions can be identified
 in the atmosphere and they are:
\begin{itemize}
    \item Region A: A Convective envelope -- where the density is
    sufficiently high such that the excess flux above the Eddington
    luminosity is advected using convection. The radiative luminosity
    left is below the classical Eddington limit: $L_{\mr{rad}} <
    \L_{\mr{Edd}} < L_{\mr{tot}}$. It was shown by Joss, Salpeter \&
    Ostriker (\jcite{JSO73}) that convection is always excited before
    the Eddington limit is reached. Thus, if the density is high
    enough and the total flux is super-Eddington, this region has to
    exist.
     
    \item Region B: A zone with lower densities, in which convection
    becomes inefficient.  Instabilities render the atmosphere
    inhomogeneous, thus facilitating the transfer of flux without
    exerting as much force.  The effective Eddington luminosity is
    larger than the classical Eddington luminosity.  $\L_{\mr{Edd}} <
    L_{\mr{rad}} = L_{\mr{tot}} < \L_{\mr{eff}}$.  $\eta$ Car has
    shown us that the existence of this region allows for the steady
    state outflow during its 20 year long eruption (\cite{etacar}).
     
    \item Region C: A region in which the effect of the
    inhomogeneities disappears and the luminosity is again
    super-Eddington.  When perturbations arising from the
    instabilities, which are expected to be of order the scale height
    in size, become transparent, the effective opacity tends to the
    microscopic value and the effective Eddington limit tends to the
    classical value.  At the transition between (B) and (C), the
    effective Eddington is equal to the total luminosity.  This
    critical point is also the sonic point in a steady state wind.
    Above the transition surface, $L_{\mr{tot}} > \L_{\mr{eff}}
    \rightarrow \L_{\mr{Edd}}$ and we have a super sonic wind.  This
    wind is expected to be optically thick.
     
    \item Region D: The photosphere and above. Since the wind is
    generally thick, the transition between regions (C) and (D) is far
    above the sonic surface.
\end{itemize}

\subsection{The location of the critical point}
\label{sec:criticalpoint}

 Paramount to the calculation of the mass loss rate in super Eddington
 systems is the location of the critical point (separating regions (b)
 and (c) in Fig.~\ref{fig:structure}), which corresponds to the sonic
 point of the wind generated at steady state. Therefore, we should try
 and estimate its location and density. Its definition is the location
 where the net {\em average} force on a gas element vanishes, or in
 other words, it is where the effective Eddington parameter
 $\Gamma_{\mathrm{eff}}$ is unity. To calculate it, we need to know
 the functional behavior of the effective opacity as a function of
 height.

 The general behavior clearly depends on the characteristics of the
 nonlinear state of the porous atmosphere.  However, this state is
 still part of an open problem under investigation and is therefore
 unknown.  It will be shown, nevertheless, that the unknown
 geometrical factors that determine the mass loss can be pin-pointed
 while some other unknowns, such as $\Gamma_{\mr{eff}}$, actually
 cancel out to first approximation.

 The effective opacity deep inside the atmosphere reduces $\Gamma$ to
 sub-Eddington values.  However, higher up in the atmosphere where the
 perturbations become optically thin, the reduction in effective
 opacity diminishes to zero.  In the thin limit, the effective opacity
 returns back to its microscopic value.  We therefore write the
 opacity (per unit mass) of the medium as:
\begin{equation}
 {\kappa_{\mr{eff}}(\Delta\tau) \over \kappa_0} =
 {\Gamma_{\mr{eff}}(\Delta\tau)\over\Gamma}
 ={\Gamma_{\mathrm{eff},\infty} \over \Gamma} +
 (1-{\Gamma_{\mathrm{eff},\infty}\over \Gamma}) \psi(\Delta\tau),
\label{eq:kappabehavior}
\end{equation}
 where $\Delta \tau$ is the equivalent optical width of a pressure
 scale height if the medium had been homogeneous.
 $\Gamma_{\mathrm{eff},\infty}$ is the effective $\Gamma$ in the limit
 of large optical depths.  It is a function of $\Gamma$ which is
 unknown at this stage since we don't have a model for the nonlinear
 behavior of the inhomogeneities.  The function
 $\Gamma_{\mr{eff}}(\Delta\tau)$ should be obtained from a non linear
 model (which is part of work in progress).  The function
 $\psi(\Delta\tau)$ gives the functional reduction of the opacity as a
 function of $\Delta\tau$.  It should depend on the geometrical
 characteristics of the inhomogeneities. For $\Delta\tau \rightarrow
 0$, $\psi \rightarrow 1$ and $\kappa_{\mr{eff}}$ will approach the
 microscopic value of the opacity $\kappa_0$.  For large optical
 depths, we have $\Delta\tau \rightarrow \infty$, and obtain
 $\psi\rightarrow 0$ and $\kappa_{\mr{eff}}\rightarrow
 (\Gamma_{\mathrm{eff},\infty}/\Gamma) \kappa_0$.

 Next, the sonic point for a steady-state transonic flow must be
 located at the critical point, where the local
 $\Gamma_{\mathrm{eff}}(\Delta\tau)$ is unity, i.e., where
 $(\kappa_{\mr{eff}}(\Delta\tau) /\kappa_0)\Gamma = 1$.  We therefore
 have:
\begin{equation}
 \psi(\Delta\tau_{\mathrm{sonic}}) = {1-\Gamma_{\mathrm{eff},\infty}
 \over \Gamma-\Gamma_{\mathrm{eff},\infty}}.
\end{equation}

 Let us assume that for small optical depths, $\psi(\Delta\tau \ll 1)
 \approx 1-\A \Delta\tau^{p_l}$ and that for large optical depths,
 $\psi(\Delta\tau \gg 1) \approx \B \Delta\tau^{-p_h}$.  If we know the
 constants $\A$ and $\B$ and the powers $p_l$ and $p_h$, we can estimate
 the average density at the sonic point.  To do that, we need a
 relation between the average density $\rho_{0}$ and {$\Delta \tau$}.
 To this goal we write the pressure scale height $l_{p}$ as
\begin{equation} l_{p}={\nu v_{s}^{2}\over
 g(1-\Gamma_{\mr{eff},\infty})}.
\end{equation} The constant $\nu$
 relates the effective speed of sound to the adiabatic one ($v_s$). If
 for example the atmosphere is isothermal, we have $\nu =
 1/\gamma$. If its temperature gradient is derived from radiation
 transfer and it has a constant opacity, we find $\nu = 3/(4 \gamma)$.
 The factor $1-\Gamma_{\mathrm{eff},\infty}$ originates from the
 `puffing-up' of the atmosphere due to the radiative force (which
 reduces the effective gravity). Let the average density be
 $\rho_{0}$, then the optical width of a pressure scale height if the
 layer is homogeneous is given by:
\begin{equation} \Delta \tau =\rho
 \kappa_{0} l_{p}={\rho \nu \kappa_{0} v_{s}^{2} \over g
 (1-\Gamma_{\mathrm{eff},\infty})},
\end{equation}
therefore:
\begin{equation} \rho = \rho_{\unit}
 \Delta\tau(1-\Gamma_{\mathrm{eff},\infty}) \quad {\mr{with}} \quad
 \rho_{\unit} \equiv g / (\kappa_0 \nu v_s^2).
\end{equation}
 
 For the small optical width limit, where $1-\psi \ll 1$, one has:
\begin{eqnarray}
 \rho_{\mathrm{sonic}} (\Delta\tau\ll 1) &\approx& \rho_\unit
 \Delta\tau_{\mathrm{sonic}} (1-\Gamma_{\mathrm{eff},\infty}) \\
 \nonumber & \approx& \rho_\unit (1-\Gamma_{\mathrm{eff},\infty})
 \left({1\over \A} {\Gamma-1 \over
 \Gamma-\Gamma_{\mathrm{eff},\infty}}\right)^{1\over p_l}.
\end{eqnarray}

 However, small $\Delta\tau$'s are obtained when $1-\psi \ll 1$ or
 $\Gamma-1 \ll \Gamma-\Gamma_{\mathrm{eff},\infty}$, namely, only when
 $\Gamma$ is very close to the Eddington limit. Under such conditions:
\begin{equation}
 \rho_{\mathrm{sonic}}(\Delta\tau\ll 1) \approx \rho_\unit {1\over 
 \A^{1\over p_l}}
 \left(\Gamma-1\right)^{1/p_l}
 (1-\Gamma_{\mathrm{eff},\infty})^{1-1/p_l}.
\end{equation}

 The opposite limit of large optical depths corresponds to $\psi \ll
 1$, and consequently to $\Gamma \gg 1$ as well.  In this limit we
 have:
\begin{equation}
 {1 \over \Delta\tau_{\mathrm{sonic}(\Delta\tau\gg 1)}} \approx
 \left({ 1- \Gamma_{\mathrm{eff},\infty} \over
 \Gamma-\Gamma_{\mathrm{eff},\infty}} {1\over \B}\right)^{1/p_h}
\end{equation}
or 
\begin{eqnarray}
 \rho_{\mathrm{sonic}} &\approx& \rho_\unit
 (1-\Gamma_{\mathrm{eff},\infty}) \Delta\tau_{\mathrm{sonic}}
 \nonumber \\&\approx& \rho_\unit \B^{1/p_h}
 \left(\Gamma-\Gamma_{\mathrm{eff},\infty}\right)^{1/p_h} \left(
 1-\Gamma_{\mathrm{eff},\infty} \right)^{1-1/p_h} \nonumber \\
 &\approx& \rho_\unit \B^{1/p_h} \left({\Gamma-1}\right)^{1/p_h} \left(
 1-\Gamma_{\mathrm{eff},\infty} \right)^{1-1/p_h}
\end{eqnarray}
 where for the last approximation we used $\Gamma \gg 1 \gtrsim
 \Gamma_{\mathrm{eff},\infty}$.

 Clearly, for the particular case of $p_h=p_l=1$ we obtain that both
 in the {\em large} and {\em small} optical depth regimes:
\begin{equation}
  \rho_{\mathrm{sonic}} \approx \rho_\unit (\A^{-1}~{\rm or}~\B) (\Gamma-1).
\label{eq:simplerhosonic}
\end{equation}
 It is interesting to note that for this particular set of power laws,
 both regimes do not depend on the value of
 $\Gamma_{\mathrm{eff},\infty}$ as it cancels out: The factor
 $1-\Gamma_{\mathrm{eff},\infty}$ introduced by the puffing up of the
 atmosphere cancels the factor $(1-\Gamma_{\mathrm{eff},\infty})^{-1}$
 which appears because when the atmosphere is closer to the effective
 Eddington limit, smaller changes in the effective opacity are needed
 to reach the critical point.

 We now proceed to show that the above simple power law scalings are
 indeed obtained under both limits. Since the arguments given are
 somewhat heuristic, the function $\psi(\Delta \tau)$ is also
 calculated analytically in the appendix for two simple cases under
 both limits, giving the same power laws. To strengthen the
 argumentation even more, we also use published results for an
 inhomogeneous system with a particular type of random statistical
 distribution (that of a Markovian statistics) for which the opacity
 (and $\psi$) can be calculated rigorously.

 To find the power laws, we assume for simplicity that the medium is
 divided into regions of higher density $\rho_h\equiv \br (1+\delta)$
 and volume $V_h\equiv \alpha V_{\mr{tot}}$ and regions of lower
 density $\rho_l\equiv \br (1-\delta)$ and volume $V_l\equiv(1-\alpha)
 V_{\mr{tot}}$.  From conservation of mass it follows that $\rho_h V_h
 + \rho_l V_l = \rho_0 V_{\mr{tot}}$ or $\br = \rho_0 / (1+(2
 \alpha-1)\delta)$.

\begin{figure}
\centerline{\epsfig{file=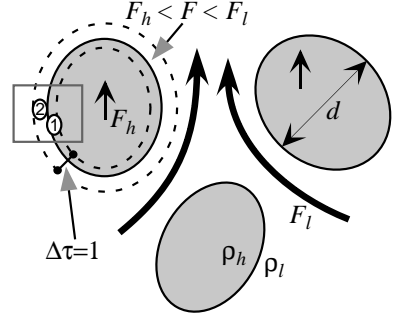,width=2.0in}}
 \caption{The effects of a finite optical depth on the flux and
 opacity of a system close to the {\em thick} limit. When the
 perturbations are optically thick, the fluxes inside and outside of
 the dense regions do not `mix'. However, because of the nonlocal
 nature of radiation transfer, a `blob' with a finite optical width
 will have a region around its interface with the low density gas in
 which the radiative flux is averaged. Part of this interface region
 (denoted by `1') has the high density, and part of it (denoted by
 `2'), has the lower density. The optical width of this region is of
 order unity. The result is that a finite fraction of the system's
 volume is not effective in reducing the opacity.}
\label{fig:thick}
\end{figure}

\vskip 8pt
\noindent
 {\bf The optically thick limit}: The optically thick limit can be
 generally described by Fig.~\ref{fig:thick}. The reduced effective
 opacity is obtained by concentrating mass in dense `blobs' and
 forming rarefied medium in between through which most of the
 radiation flux is funneled through. Simplistically, we have high
 density regions with a volume $V_h$, a density $\rho_h$ and a reduced
 flux $F_h$ and low density regions with $V_l$, $\rho_l$ and higher
 typical fluxes $F_l$.

 In the absence of inhomogeneities, we would have had an average
 density $\rho_{0}$ and a flux $F_{0}$. For a given temperature
 gradient, we typically have $F_{h,l} \approx F_{0}
 \rho_{0}/\rho_{h,l}$. The result is accurate if the perturbations in
 the medium have a large vertical to horizontal aspect ratio, such
 that the temperature perturbations are negligible. Under more general
 perturbation geometries, this approximation is only a rough estimate
 but it should give the typical change expected in the fluxes.

 The total average force $f$ acting on a volume which contains many
 inhomogeneous elements, is $f = \kappa_0( V_h \rho_h F_h + V_l \rho_l
 F_l) \approx (V_{l}+V_h) \kappa_0 \rho_{0} F_{0}=f_{0}$, where
 $f_{0}$ is the force corresponding to the homogeneous case.  That is,
 the average force does not change for a given fixed average
 temperature gradient.  But the flux through the system is
\begin{eqnarray}
 {F \over F_0}&=& {V_h F_h + V_l F_l\over F_0 V_{\mr{tot}}} \approx
 \left({{V_h \over \rho_h} + {V_l \over \rho_l}}\right) {\rho_{0}\over
 V_{\mr{tot}}} \nonumber \\ &=& {1-\left[(2 \alpha-1)\delta\right]^2 \over
 1-\delta^2} \equiv (\kappa_0/\kappa_{\mathrm{eff}})  \geq 1.
\label{eq:F0}
\end{eqnarray}
  We find an increased flux without changing the driving temperature
  gradient.  Said in other words, the same temperature gradient drives
  now a higher flux---the radiative conductivity increases or the
  effective opacity decreases.

 What happens though when the optical depth of the perturbations is
 finite?  If we look at Fig.~\ref{fig:thick}, we see that there exists
 a finite volume at the interface between the high density regions and
 the low density ones with a total cross section of about one optical
 depth.  The material in this volume sees radiation from both the high
 density regions and the low density regions, so its flux will be
 averaged $F_{\mr{avr}}\approx (F_h+F_l)/2$.  The amount of mass that
 witnesses this averaged flux is of order the surface area $S$ of the
 interface between high and low density regions times $\br
 \kappa_{0}$.  Half of the mass will be on the dense side (labeled
 `1' in the figure) and occupy a volume of order $V_1 \approx S / 2
 \rho_h \kappa_0$ and the other half on the rarefied side (`2') and
 will occupy a larger volume $V_2 \approx S / 2 \rho_l \kappa_0$.
 When calculated, the total force $f_{\mr{tot}}$ for a given
 temperature gradient remains unchanged, but the flux will not be as
 high:
\begin{eqnarray}
 {F\over F_0} \hskip -3pt &\approx& \hskip -3pt {\rho_0 \over
 V_{\mr{tot}}} \hskip -1pt \left[ {V_h - V_1 \over \rho_h}+{V_l - V_2
 \over \rho_l} + {V_1 + V_2 \over 2}
 \left({1\over\rho_h}+{1\over\rho_l} \right) \right] \nonumber \\
 &\approx& \hskip -3pt {F_{tot,\Delta\tau \rightarrow \infty}\over
 F_0} - {(V_2 - V_1)(\rho_h - \rho_l) \rho_0 \over 2 V_{\mr{tot}}
 \rho_h \rho_l} \nonumber \\ \hskip -3pt &\approx& \hskip -3pt
 {\kappa_0\over \kappa_{\mr{eff},\infty}} - {S \delta^2 (1+(2 \alpha
 -1)\delta)^2 \over (1-\delta^2)^2 \kappa_0 \rho_0 V_{\mr{tot}}}
\end{eqnarray}

If we further assume that $\Gamma \gg 1$, then the significantly
under-dense regions carry the lion share of the flux while the denser
regions transport only a minor fraction.  Under such conditions one
can neglect $1/\rho_h$ when compared with $1/\rho_l$ or assume that
$\delta$ is close to unity.  We then obtain:
\begin{equation}
 {F\over F_{0}} \approx {\kappa_0 \over
 \kappa_{\mathrm{eff},\infty}} - \left( {\kappa_0 \over
 \kappa_{\mathrm{eff},\infty}} \right)^2 {S \over \kappa_0 \rho_{0} 4
 (1-\alpha)^2 V_{\mr{tot}}} .
\label{eq:Fthick}
\end{equation}

The effective opacity is then:
\begin{equation}
 {\kappa_{\mathrm{eff}}\over \kappa_0} = {f \over f_{0}}{F_{avr} \over
 F_{\mr{tot}}} \approx {\kappa_{\mathrm{eff},\infty} \over \kappa_0} +
 {S \over 4 (1-\alpha)^2\kappa_0 \rho_{0} V_{\mr{tot}}}.
\end{equation}
 The more elaborate treatment in the appendix gives that the prefactor
 (again, for vertically elongated perturbations) is not
 $1/(4(\alpha-1)^2)$ but $3/(32\alpha(\alpha-1)^2)$ instead (i.e., 3/4
 times smaller for $\alpha=1/2$).

 The term $(S / 2 \kappa_0 \rho_{0} V_{\mr{tot}})$ can be related to
 the optical depth of the perturbations.  If the physical size of the
 perturbations is $d$, then we can write $S = \Xi V_{\mr{tot}} / d$,
 where $\Xi$ is an unknown geometrical factor.  The geometrical factor
 is equal for example to 1 if the geometry is that of sheets of width
 $d$.  It is higher for higher dimensionally of the structure (e.g.,
 $\Xi \approx 2~ {\rm or}~ 3$ for cylinders or spheres of diameter
 $d$, if $\alpha = 1/2$).

 If we now compare this result with the definition of the function
 $\psi$, we have:
\begin{equation}
 1 - \psi
 \approx { 3 S \over 32 \alpha ( 1-\alpha)^2 \kappa_0 \rho_0
 V_{\mr{tot}}} \approx \left( 3 \Xi \over 32 \alpha (1-\alpha)^2
 \right) {1 \over \beta}{1\over \Delta\tau},
\end{equation}
 where $\beta \equiv d/l_{p}$. It is the ratio between the typical
 size of the perturbations and the pressure scale height of the
 atmosphere. Since the typical wavelengths which are the quickest to
 grow are those with a size comparable to that of the scale height
 (\cite{instabilities}), this factor should too be of order unity but
 could be somewhat larger or smaller, depending on the details.  The
 expression in parenthesis describes a geometrical factor which is of
 order unity or somewhat larger (e.g., if $\alpha=1/2$ then it is
 about 1-3 for typical $\Xi$'s, and it would typically be larger for
 other $\alpha$'s).

 Through the definition of $\psi$ in the thick limit, we have
\be
 \B= { 3 \Xi \over 32 \alpha \beta (1-\alpha)^2} \quad {\mr{and}} \quad
 p_h = 1.
\label{eq:B}
\ee

 \vskip 8pt \noindent {\bf The optically thin limit}: We assume for
 simplicity that the medium is made up of spherical regions of higher
 density and those with lower density, each occupying a similar
 volume.  Under this approximation, we can estimate both $\A$ and $p_l$
 without cumbersome algebra.

 The equation of radiative transfer for a gray atmosphere is:
\begin{equation}
{dI(\vx)\over d s} = \chi(\vx)( B(\vx) -  I(\vx)),
\label{eq:radeq}
\end{equation}
 where $\chi = \kappa \rho$ and $B$ are the opacity per unit volume
 (i.e., the extinction) and radiation source function respectively.

 The formal solution to the radiative equation for a point $\vx$
 located many optical distances from the boundary is (e.g.,
\cite{MWM}):
\begin{equation}
 I(\vx,\hn) = \int_0^\infty \chi(\vx-\hn s) B(\vx-\hn s)
e^{-\int_0^s \chi(\vx-\hn s')ds'} ds.
\label{eq:formalsol}
\end{equation}

\begin{figure}
\centerline{\epsfig{file=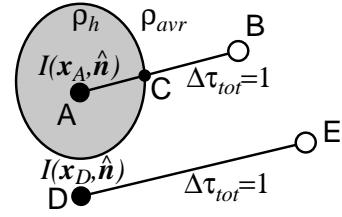,width=1.75in}}
 \caption{The effects of a finite optical depth on the flux and
 opacity of a system close to the {\em thin} limit. Point {\sf A}
 located in a higher density region will witness a slightly reduced
 flux from the average one. The main reason is that when calculating
 the intensities $I({\bf x}_A,\mu)$ at point {\sf A}, the physical
 distance to a point {\sf B} located an optical distance of unity from
 point {\sf A} will be smaller than the distance {\sf D-E} between two
 points with an average density between them, because the path {\sf
 A-C} has a higher density. This short path will affect all photons
 coming from outside the dense region, giving a linear correction in
 $\rho_0 \kappa_0 d$. Because $I({\bf x}_A,\mu)$ are sampled from a
 smaller volume, the first moment of $I$ which corresponds to the
 flux, will be smaller. }
\label{fig:thin}
\end{figure}

 Next, we calculate the intensities $I(\vx_A,{\hn})$ at point `{\sf A}'
 located inside a spherical region of radius $r_0$ with a density
 $\rho_0(1+\delta)$ (cf fig.~\ref{fig:thin}).  We assume that outside
 the given high density region the photons seen many high and low
 density regions and hence an average density.  In other words, there
 is a high density correlation inside a given blob but it is assumed
 to be negligible for distances larger than the blob.

 We also assume that the radiation source function is $B=B_0+b z$ and
 that there is no contribution from the perturbations.  In the
 appendix, it will be shown to be a valid assumption because the
 correction to the source function is of second order, i.e., it falls
 as $(r_0 \chi_0)^2$.  We also neglect higher order terms in $z$.  We
 define now $\mu$ as the angle between the integration path and $\hz$. 
 For simplicity, we also assume that $\alpha=1/2$.

 Using the formal solution, the radiation intensity is then given by:
\begin{eqnarray}
 I(\vx_A,{\hn}) \hskip -3pt &= \hskip -3pt & B_0 + \chi_0 \cos(\mu) \\
 && \times \left[ \int_0^{r_0} \chi_{0} s (1+\delta) \exp(-\chi_0 s
 (1+\delta)) ds \right. \nonumber \\ \nonumber && \left. +
 \int_{r_0}^{\infty} s \exp(-\chi_0 \rho_0 ((1+\delta) r_0 + (s-r_0))
 \right] \\ &\approx \hskip -3pt& B_0 + {b\over \chi_0} (1- \delta r_0
 \chi_0) \cos \mu +{\cal O}(r_0 \chi_0)^2.
\end{eqnarray}

 The flux is then given by:
\begin{equation}
 F_h = F_A = {1 \over c} \oint I(\vx_A,{\hn}){\hn} d\Omega = F_0 (1-\delta)
\end{equation}
 where $F_0 = 4 \pi b/(3 c \chi_0)$.

 The flux in the under-dense regions is obtained by the transformation
 $\delta\rightarrow -\delta$. In the simplest case discussed here of
 both regions having the same volume fraction, we see that the total
 flux through the system remains unchanged. However, the mean force is
 now reduced:
\begin{equation}
f = V_h \rho_h \kappa_0 F_h + V_l \rho_l \kappa_0 F_l = f_0 (1- r_0
\kappa_0 \rho_{0} \delta^2).
\end{equation}
 The effective opacity is therefore reduced as well. Using the result
 of eq.~(\ref{eq:F0}) for the thick limit and $\alpha =1/2$, and
 comparing to eq.~(\ref{eq:kappabehavior}) with $\psi(\Delta\tau \ll
 1)\approx 1-\A\Delta\tau^{p_l}$, we finally get
\be
 \A= {\kappa_0 \rho_0 d \over 2 \Delta \tau} ={d\over 2 l_{p}}=\beta/2 \quad
 {\mr and} \quad p_l=1,
\ee
 with $d=2 r_0$.  In other words, also in the thin limit, the function
 $\psi$ depends only on geometrical factors (and not for example on
 $\Gamma_{\mr{eff},\infty}$).

\vskip 8pt
\noindent
 {\bf A Markovian mixture}: A special case which can be solved more
 rigorously for any optical thickness is that of a Markovian mixture
 of dense and rarefied phases. A Markovian mixture is defined through
 the spatial statistics of the distribution: On any ray passing
 through the medium, the interface between material (or phase) A and
 material B form a Poisson process. In other words, if at point ${\bf
 s}$ the fluid is of type A, then at an infinitesimally close point
 ${\bf s}+d{\bf s}$, the probability of finding fluid B is
 $ds/\lambda_0$. If point ${\bf s}$ is in fluid B, then the equivalent
 probability of finding fluid A at point ${\bf s}+d{\bf s}$ is
 $ds/\lambda_1$. The effective transport properties in such a mixture
 were extensively analyzed by \cite{Levermore86} and
 \cite{Levermore88}, and it is described in detail in \S3.4 of
 \cite{Pomraning92}. It was found that the effective absorption
 cross-section in the absence of scattering is:
\begin{equation}
 \sigma_{\mathrm{eff}} =
 {(p_0\sigma_0+p_1\sigma_1+\sigma_0\sigma_1\lambda_c)\over
 (1+(p_0\sigma_1+p_1\sigma_0)\lambda_c)},
\end{equation}
 with $\lambda_c \equiv {\lambda_0\lambda_1 / (\lambda_0+\lambda_1)}$,
 $p_i \equiv \lambda_i/(\lambda_0+\lambda_1)$ and $\sigma_{0,1}$ being
 the two absoption cross-sections of materials $A$ and $B$
 respectively. To translate this result to our problem notation, we
 identify $p_0$ with our $\alpha$, $p_1$ with $1-\alpha$ and
 $\lambda_0$ with $d$, such that $\lambda_1$ is $d (1-\alpha)/\alpha$
 and $\lambda_c = (1-\alpha) d$. Moreover, we can identify $\sigma_0$
 with $\rho_h \kappa_0$ and $\sigma_1$ with $\rho_l \kappa_0$. We then
 find that the effective opacity at the large optical depth limit is:
\begin{equation}
 {\kappa_{\mathrm{eff},\infty} \over \kappa_0} = {1-\delta^2 \over
 (1+(2 \alpha-1)\delta)^2 }.
\end{equation}
Next, we write $d = \beta l_p$ and $\Delta \tau = l_p \rho_0
\kappa_0$.  Then, if we further assume that this reduction is large,
as we did before, we can assume that $\rho_h \gg \rho_l$ or $\delta
\approx 1$. Using this result, in the two optical depth limits
($\Delta\tau \ll 1$ for $\cal A$ and $\Delta \tau \gg 1$ for $\cal B$), it
is straightforward to show that:
\begin{equation}
 {\cal A} = (1-\alpha) \beta \quad {\rm and} \quad {\cal B} = {1 \over
 (1-\alpha) \beta},
\label{eq:Markovian_res}
\end{equation}
 as well as $p_h=p_l=1$. Namely, in the special case of a Markovian
 mixture, we find that the function $\psi(\Delta\tau)$ behaves in the
 same way as in the simple models presented before, with only the
 normalization constants $\cal A$ and $\cal B$ being somewhat
 different but of the same order as before.

 \vskip 8pt Although the analysis carried out here for the location of
 the critical point in both limits is far from accurate, it shows us
 that a simple formula (given by eq.~\ref{eq:simplerhosonic}) should
 describe the average density at the sonic point.  Moreover, in both
 limits, the unknown constants depend only on geometrical factors such
 as the size of the perturbations in units of the pressure scale
 height ($\beta$) or the ratio $\Xi$ between the interface surface
 between high and low density regions and their volume in units of
 $d$, as well as on the particular shape the inhomogeneities adopt.

 In both limits, the unknown constants $\A^{-1}$ and $\B$ were shown to
 be proportional to $\Delta\tau / \kappa_0 \rho_0 d=\beta^{-1}$.
 Since the perturbations are expected to be of the order of the scale
 height, this factor is of order unity as well.  If the perturbations
 are smaller, the term will be larger and with it the density.

 We define now a wind function ${\cal W}(\Gamma)$ such that:
\begin{equation}
 \rho_{\mr{sonic}} \equiv \sqrt{\nu} {\cal W}(\Gamma) \rho_\unit
 (\Gamma-1)
\label{eq:Wrhosonic}
\end{equation}
 (the factor $\sqrt{\nu}$ is introduced so that the ensuing wind
 mass-loss relation will not include it).  From our analysis thus far,
 we found that if the geometrical properties of the developed
 inhomogeneities are not a function of $\Gamma$, then the prefactor to
 $(\Gamma-1)$ should be constant (but might be different for the two
 regimes: $\Gamma \gg 1$ and $\Gamma-1 \ll 1$).

 Since  we have no information yet as to if and how the geometrical
 properties change with $\Gamma$, we {\em assume} in the remainder of
 the paper that they are constant and that the wind function is a
 constant as well. Thus,
\be
 \func(\Gamma) = {1\over \sqrt{\nu}}(\A^{-1}~{\mr{or}}~\B) \approx
 const.
\label{eq:WAB}
\ee

 We should emphasize that a simplifying assumption that we implicitly
 made is that the same proportionality relation exists for the large
 and the small optical depths. This is found to be the case in the
 Markovian mixture of low and high density regions. However, this need
 not be the general case.

 We should also bare in mind that the transition between the two
 normalizations (i.e., a different ${\cal W}$ for the optically thick
 and thin cases) should take place at $\Gamma - 1 \sim
 1-\Gamma_{\mathrm{eff},\infty} \lesssim 1-\Gamma_{\mr{crit}} \approx
 0.15 - 0.5$ (depending on the type of instability that governs). This
 will actually be below the typical $\Gamma$'s we will obtain in the
 analysis. That is, we are mainly concerned with the optically thick
 limit.  We should also be cautious because the function $\psi(\Delta
 \tau)$ could in principle also be a function of $\Gamma$ (and
 therefore the constants $\A$ and $\B$ as well), especially if the
 geometrical behavior of the inhomogeneities changes with
 $\Gamma$. For example, if the typical optical depth of the `blobs' is
 $\Gamma$ dependent.  Only a detailed numerical analysis could shed
 more light on this problem.

 The next step in the analysis presented here is to follow the
 nonlinear characteristics of the inhomogeneities.  This is an
 intrinsically difficult problem since it requires the
 radiation-hydrodynamical understanding of the saturation properties
 of the instabilities in the medium.  This analysis has commenced in
 the form of a detailed hydrodynamic simulation.  A different approach
 which will soon be completed, is to find empirically ${\cal W}$ and
 any possible $\Gamma$ dependence.  This can be performed by solving
 for the super-Eddington counterpart to the core-mass luminosity
 relation (\cite{Paczynski70}) which should describe the steady state
 of novae in their post-maximum decline.  Then, by comparing to novae,
 both $\Gamma_{\mathrm{eff},\infty}$ and ${\cal W}$ can be extracted
 for novae with different luminosities.

\subsection{The mass loss rate}

 Once we have estimated the density at the sonic point, we can proceed
 to calculate the mass loss rate. It is given by: $\Md=4 \pi R_0^2
 \rho_{\mr{sonic}} v_{\mr{sonic}}$, where the relevant speed of sound
 is $v_{\mr{sonic}} = \sqrt{\nu} v_s$. This speed is generally not the
 adiabatic speed of sound since the flow isn't adiabatic.

 Using eq.~(\ref{eq:Wrhosonic}), we find:
\ba
 \Md &=& 4\pi R_\st^2 \rho_{\sonic} v_{\sonic} = { 4 \pi G M_\star
 \func(\Gamma) (\Gamma-1) \over \kappa_0 v_s} \nonumber \\ &=&
 {\func(\Gamma) (L-\L_{\mr{Edd}}) \over c v_s}.
\label{eq:masslossrate}
\ea

 Note that $\L_{\mr{Edd}}$ becomes the modified Eddington
 luminosity if the underlying opacity is not that of Thomson
 scattering.  Eq.~(\ref{eq:masslossrate}) is the basic result of the
 present theory.
 
 From the analysis of the location of the critical point in
 \S\ref{sec:criticalpoint}, we have seen that $\func$ should be of
 order unity or somewhat larger if the size of the perturbations $d$
 is similar to the pressure scale height of the atmosphere (since
 $\cal W$ scales with $l_p/d$).  Moreover, since $\cal W$ depends only
 on the geometrical properties of the perturbations, it could be close
 to a constant if these properties are not a function of $\Gamma$.
 Hence, we assume for simplicity that $\func(\Gamma)$ is a constant of
 order unity.  We iterate once more that it is only with a more
 elaborate simulation or with more accurate and careful observations,
 that a more accurate functional form can be deduced.  For the
 meantime, we will have to settle with this simplifying yet reasonable
 assumption.

 We will show in the rest of the paper that
 eq.~(\ref{eq:masslossrate}) provides an explanation to the mass loss
 from bright classical novae as well as from $\eta$ Carinae and allows
 us to connect between the observed luminosity and mass loss rate, a
 relation which hitherto did not exist for super-Eddington systems.

\subsection{High Load winds}
\label{sec:load}

 Unlike normal stellar winds, the thick winds formed in the
 super-Eddington flows not only have a high mass momentum relative to
 the total radiative momentum, which is expected in any thick wind,
 but more importantly, the kinetic energy flux can be comparable to
 the radiative luminosity.  It also inadvertently implies that the
 rate of gravitational energy `pumped' into the outflowing matter is
 also comparable.  Consequently, the luminosity $L$ in
 eq.~(\ref{eq:masslossrate}) is not the observed luminosity at
 infinity.  Instead, we have to substitute $L_{0}$ for $L$ where
\begin{eqnarray}
 L_\st \approx L_{\mr{tot}}& =& L_\infty + \Md \left({v_\infty^2 \over 2}
  + {GM_\star \over r_\st}\right) \nonumber \\ &=& L_\infty + {\Md \over 2}
 (v_\infty^2 +v_{\mr{esc}}^2).
\end{eqnarray}
 (The approximation neglects the kinetic energy at the sonic points,
 i.e., assuming that $v_s^2 \ll v_{\mr{esc}}^2$ which gives
 $L_{\mr{tot}} \approx L_\st$).  The effect of a lowered observed
 luminosity was coined `photon tiring' by Owocki \& Gayley
 (\jcite{Owocki97}), who solved for the behaviour and evolution of the
 wind.  Their solution related the variables $L_{\infty}, L_{\st},
 \Md, v_\infty^2$ and $v_{\mr{esc}}^2$.  The basic equations are the
 equations of continuity, momentum conservation and energy
 conservations: \ba \Md &=& const = 4 \pi r^2 v \rho \\ {1\over 2} {d
 v^2 \over d r } &=& - {G M_\star (1-\Gamma(r)) \over r^2} \\ L(r) &=&
 L_\st - \Md \left[{v^2 \over 2} + {GM_\star\over R_\st} -
 {GM_\star\over r} \right], \ea where $\Gamma(r)$ is the ratio between
 the luminosity $L(r)$ and the local modified Eddington luminosity
 $\L_{\mr{Edd}}$ which we assumed to be constant, since the opacity is
 assumed to be given by the Thomson scattering.  The solution to this
 set of equations, after neglecting the speed of sound at the base of
 the wind, is (\cite{Owocki97}): \be w \equiv {v^2\over
 v_{\mr{esc}}^2} = - x + {\Gamma_0 \over \tm} \left(1-\exp(-\tm
 x)\right) , \ee where $x$ is defined as $1-{R_\st/ r}$.  $R_{\st}$ is
 the radius of the sonic point of the wind and can be described as the
 `hydrostatic' radius of the star.  Below it, the envelope is
 expanding with a subsonic speed.  $\tm$ is defined as $\Md G M_\star
 / \L_{\mr{Edd}} R_\st$.  It is a slightly different definition for
 the `photon tiring number' than the original definition of $m$ by
 Owocki \& Gayley (\jcite{Owocki97}).  The two are related through
 $\tm = \Gamma_0 m$.

 At very large radii ($r\rightarrow\infty$), we have
\be
 w_\infty = {v_\infty^2 \over v_{\mr{esc}}^2}= - 1 + {\Gamma_0 \over \tm}
 \left(1-\exp(-\tm)\right).
\label{eq:vel-inf}
\ee

 Using the wind model developed here (eq.~\ref{eq:masslossrate}), we
 can close the relation between $\Md$ (or $\tm$) and the luminosity
 ($\Gamma_{0}{\cal L}_{\mr{Edd}}$) of the star:
\be
 \tm = {1\over 2}{v_{\mr{esc}}^2 \Md \over \L_{\mr{Edd}} } = {1 \over
 2}{v_{\mr{esc}}^2 \func \over v_s c} (\Gamma_0 -1) \equiv (\Gamma_0 -
 1) \Wt,
\label{eq:scaledmassloss}
\ee
 where $\Wt$ is the `scaled' wind parameter. This is a dimensionless
 version of eq.~(\ref{eq:masslossrate}) -- the basic mass-loss
 luminosity relation of SEWs.

 For given $\Gamma_0$, $\L_{\mr{Edd}}$ and $R_\st$ (or
 $v_{\mr{esc}}^2$), we can now calculate $\tm$ (or $\Md$), and from it
 calculate $w_\infty$ (or $v_{\infty}^2$) and $\Gamma_\infty$. To get
 the latter, we substitute the energy conservation equation into the
 equation for $w$ to get:
\be
\Gamma_\infty = \Gamma_0 \exp(-\tm).
\label{eq:gamma-inf}
\ee
 In order to compare with observations, we need to translate
 $L_\infty$, $\Md$ and $v_\infty$, into an observed temperature.  This
 is done using a steady state optically thick wind model. When
 integrated upwards, a photosphere should be obtained where the
 optical depth is of the order of 2/3\footnote{The photosphere in a
 spherical geometry does not sit at exactly $\tau=2/3$, though we
 assume so for simplicity.}. The general case is far from trivial
 because the opacity is $\sim\sqrt{\kappa_{\mr{sc}} \kappa_{\mr{ab}}}$
 where $\kappa_{\mr{ab}}$ is the absorptive opacity which is generally
 much smaller than the scattering opacity.  The simplest approximation
 is first to assume that the opacity is that of Thomson
 scattering. This was done when the original steady state optically
 thick wind explanation to novae was proposed (\cite{Bath76}). It
 yields an effective temperature that satisfies the following
 equation:
\ba
\label{eq:BSTeff}
 \log \left(\Md \over \mr{gr~s^{-1}}\right) &=& 22.0 - 2.0 \log
 \left(T_{\mr{eff}} \over 10^{4}~\mr{{}^{\circ}K}\right) \\
 & & \hskip -1.2cm + 
 \log\left(v\over10^8 \mr{~cm~s^{-1}}\right) + 0.5
 \log \left( L_{\infty} \over 10^{38} \mr{~erg~s^{-1}} \right).
 \nonumber
\ea
 In reality though, the absorption opacity is lower and the
 photosphere is deeper in the wind, where the temperature is
 higher. Thus, a temperature estimate based on eq.~(\ref{eq:BSTeff})
 will be adequate only when we use an effective temperature that is
 defined through $\sigma T_{\mr{eff}}^4 \equiv L/ 4 \pi
 r_{\mr{ph}}^2$, where $r_{\mr{ph}}$ is the radius where the optical
 depth for the continuum becomes $2/3$. This temperature is used, for
 example, in the analysis by Schwarz et al.~(\jcite{Schwarz98}). If a
 real colour temperature is needed (such as by comparison to a
 Planckian spectrum), then a more extended analysis, as was carried
 out by Bath (\jcite{Bath78}) is needed. Since we do not require a
 very accurate mass loss temperature relation (as the measurement
 error in the observations is much larger anyway), a simple fit to
 Bath's results is sufficient. One obtains that for
 $T_{\mr{colour}}\gtrsim 6000^\circ{\mr{K}}$, the relation is:
\ba
\label{eq:BSTcolour}
 \log \left(\Md \over \mr{gr~s^{-1}}\right) &=& 22.3 - 1.48 \log
 \left(T_{\mr{colour}}\over 10^4\mr{{}^\circ K}\right) \\ & & \hskip -1.4cm
 + \log\left(v\over10^8 \mr{~cm~s^{-1}}\right) +0.65 \log \left(
 L_\infty \over 10^{38} \mr{~erg~s^{-1}} \right).
\nonumber
\ea
 It provides an estimate to $\log \Md$ that is better than $0.1$,
 which is more than sufficient for our purpose here.

 Both eqs.~(\ref{eq:BSTeff}) \& (\ref{eq:BSTcolour}) can be written
 more generally as
\be
 {\Md \over \Mdotbar} = {(v_\infty/\vinfbar)(L_\infty/\Lbar)^{p_L}
 \over (T_i/\Tbar)^{p_T}},
\label{eq:BSgeneral}
\ee
 where $T_i$ is either $T_{\mr{eff}}$ or $T_{\mr{colour}}$ and
 $\Mdotbar, \vinfbar,\Lbar,\Tbar, p_L $ and $p_{T}$ are constants read
 off eqs.~(\ref{eq:BSTeff}) \& (\ref{eq:BSTcolour}). This expression
 allows us to treat both cases simultaneously.

 The last relation that we need in order to close the set of equations
 is the exact value of $v_s$ at the base of the wind.  To this goal we
 need the temperature at the base of the wind.  For simplicity, we
 assume that the velocity is constant and equal to the terminal value;
 in other words, most of the acceleration takes place deep in the
 optically thick part.  This assumption is valid as long as
 $r_{\mr{ph}}\gg R_{\st}$.  Under these conditions, the temperature at
 the bottom can be estimated to be: \be T_\st \approx T_{\mr{i}}
 \left(r_{\mr{ph}} \over R_\st\right)^{3/4}
 \left(\Gamma_0\over\Gamma_\infty \right)^{1/4}.
\label{eq:T-sound}
\ee
 When we work with the colour temperature, $r_{\mr{ph}}$ is the depth
 of the last mean emission (i.e., $\tau=2/3$ for an opacity of
 $\sqrt{\kappa_{\mr{sc}} \kappa_{\mr{ab}}}$).

 On one hand, the mass loss depends on the parameters of the sonic
 point (e.g., $R_0$ or $v_{\mr{esc}}$, and $T_0$).  On the other hand
 however, these parameters depend on the photospheric conditions
 (e.g., through eq.~\ref{eq:T-sound}).  Therefore, the parameters of
 the wind, the conditions at the sonic point and the conditions at the
 photosphere should all be solved self consistently.  It is carried
 out as follows:

 Assume that the mass of an object $M_\star$, its Eddington parameter
 at the base $\Gamma_0$, and a scaled wind parameter $\Wt$ are given.
 To obtain the complete wind, sonic point and photospheric parameters
 we begin by calculating $\tm$ using
 eq.~(\ref{eq:scaledmassloss}). Next, $\Gamma_\infty$ (and $L_\infty$)
 are found using eq.~(\ref{eq:gamma-inf}). $w_\infty$ is then obtained
 using eq.~(\ref{eq:vel-inf}). To obtain the remaining parameters, we
 have to combine together the rest of the equations. These are
 eqs.~(\ref{eq:scaledmassloss}), (\ref{eq:T-sound}), and
 (\ref{eq:BSgeneral}). The latter corresponds to either
 eq.~(\ref{eq:BSTeff}) or eq.~(\ref{eq:BSTcolour}), depending on
 whether we work with the effective temperature $T_{\mr{eff}}$ or the
 colour temperature $T_{\mr{colour}}$. To close the relations we need
 three more equations, which are $v_{\mr{esc}}^2 = 2 GM_{\star}/R_0$,
 $v_s = \sqrt{\gamma kT_0/\mu m_H} \equiv \vsbar \sqrt{T_0/\Tbar}$ and
 $\sigma T_i^4 = L_\infty / 4 \pi r_{ph}^2$ .  When combined, we find
 the following expressions for $T_i$, $v_\infty$ and the mass loss
 $\Md$:
\ba
\nonumber 
   T_i &\bnsk=\bnsk& \nsk \left[c^{12}  \Ledd^{5p_L-{11\over4}}
   \Mdotbar^5 \Tbar^{5p_T-6} \vsbar^{12}w^{5\over2}\Wt^{12}
   \Gamma_0^{3\over2} \Gamma_\infty^{{3\over4}+5p_L} \over 4
   \pi^{9\over 4} \sigma^{9\over 4} G^{9\over2} \Lbar^{5p_L} \tm^5M_{\star}^{9/2}
   \vinfbar^5 \func^{12} \right]^{1\over 3+5 p_T}
\\ \nonumber 
   v_\infty &\bnsk=\bnsk& \nsk \left[{2 \Lbar^{p_L} \tm \vinfbar w
   \over \Gamma_\infty^{p_L} \Ledd^{p_L-1} \Mdotbar } \left( 2 c^{4}
   \vsbar^{4} \Wt^{4} \Gamma_0^{1\over2}
   \Gamma_\infty^{1\over4}\Ledd^{3\over 4} w^{5\over 2} \over
   \pi^{3\over 4} \sigma^{3\over 4} G^{3 \over 2} M_\star^{3 \over 2}
   \Tbar^{3 } \func^{4} \right)^{p_T} \right]^{1\over 3+5 p_T}
\\ \nonumber 
 \Md &\bnsk=\bnsk& \nsk \left[ { 2 \Ledd^{1+2p_L} \tm \Mdotbar^2 w \Gamma_\infty^{2
 p_L} \over \Lbar^{2 p_L} } \right. \\ &\bnsk\bnsk & \left. \times \left( 8 \pi^{3\over 8} G^3
 \Ledd^{7\over2} \tm^5 M_\star^3 \Tbar^6 \func^8 \sigma^{3\over2} \over c^8
 \vsbar^8 \Wt^8 \Gamma_0 \Gamma_\infty^{1\over 2} \right)^{p_T}
 \right]^{1\over 3+5 p_T}.
\ea
 The expressions for the  additional parameters can be found in a
 straightforward manner.

\begin{figure} 
\centerline{\epsfig{file=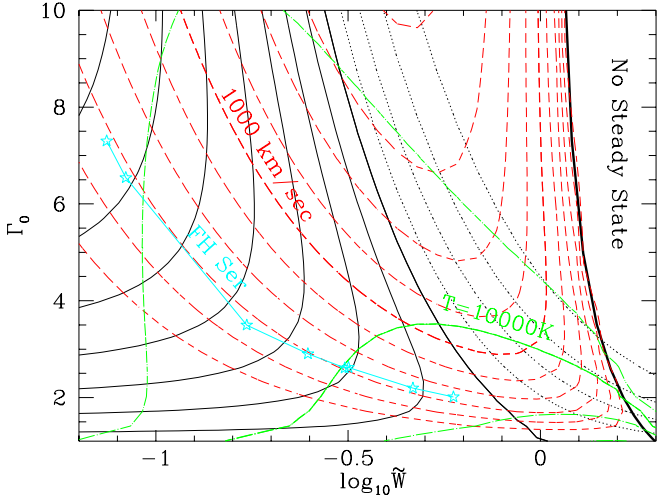,angle=-90,width=\figurewidth}}
\caption{
  {The state solution in the $\Gamma_0$ - $\Wt$ plane.  Here,
  $\Gamma_{0}$ is the Eddington parameter at the base of the wind,
  while $\Wt$ is the scaled wind parameter describing the relative
  `load' of the wind $\Wt \equiv W v_{\mr{esc}}^2 / 2 v_s c$.  Larger
  $\Wt$'s imply a smaller base radius.
  The thin continuous set of lines is lines of constant
  $\Gamma_{\infty} > 1$. The separation between two consecutive lines
  is 0.2 dex.  The set of dotted lines is lines of constant
  $\Gamma_{\infty} < 1$ separated by 0.2. The $\Gamma_{\infty}=1$ line
  separating between the two sets of lines, is marked heavier.
  The dashed set of lines is lines of constant $v_{\infty}$ separated
  by $100$ km s$^{-1}$. The asymptotic line of $v_{\infty}=0$, beyond
  which there is no steady state wind solution, is marked as a thick
  line.
  The dot-dash set of lines describe constant colour temperatures 
  separated by 2500K. Note that the lower lines are the hotter ones.
  The evolution of FH Ser is marked by stars connected with a
  continuous line.  The evolution of FH Ser from top left to bottom
  right was calculated by combining the following equations as
  explained in the text: (a) The wind model to relate $\Md$ to
  $\Gamma_0$ (given in this work), (b) The temperature mass loss
  relation (given by Bath 1978), and (c) Relations between the
  observed parameters and the parameters at the base of the wind
  (given by Owocki \& Gayley 1997). The stars are the values obtained
  at different days from the data of Friedjung (1987). The fact that
  the light curve conspicuously follows an iso-$v_\infty$ contour
  arises because the observed $v_\infty$ for FH Serpentis happens to
  vary only little, from 670 to 770 km s$^{-1}$.}  }
\label{fig:contours}
\end{figure}

 The contour plots of $\Gamma_\infty(\Gamma_0,\Wt)$,
 $v_\infty(\Gamma_0,\Wt)$ and $T_{\mr{colour}}(\Gamma_0,\Wt)$ are
 depicted in Fig.~\ref{fig:contours} for a $1 M_{\sun}$ system.

\section{Application of the SEW theory}
\label{sec:novae}

We now proceed to apply the SEW theory to systems that clearly
exhibited super-Eddington outflows over long durations relative to 
the dynamical time scales.

\subsection{Two super-Eddington nova}

 Although many novae display super-Eddington luminosities for at least
 short periods, it is often difficult to unequivocally show that a
 particular nova was indeed super-Eddington for a {\em long} duration,
 even if it was\footnote{On average, the absolute magnitude $M_{B}$ of
 a classical novae of which the WD mass $ \gtrsim 0.5 M_{{\sun}}$,
 peaks at a super-Eddington luminosity (\cite{Livio92}).}. The reason
 is that during the `bolometric plateau' that a nova exhibits, the
 luminosity can be close to the Eddington limit (from either above or
 below it) such that even small uncertainties in distance and
 luminosity can hide the true nature of the flow. Moreover, since the
 wavelength of maximum emission shifts to the UV during the
 `bolometric plateau', as the temperature increases, it is difficult
 to acquire accurate bolometric luminosities from Earth based
 observations.  Our insistence of using good data results with having
 only two clear cases in which long duration super-Eddington flows
 were observed and have good enough data. The two cases are \NLMC~and
 \NFH.  Other cases such as V1500 Cygni are not as clear cut, even
 though one would suspect that long duration SEWs could have been
 present. In V1500 Cygni, the nova was observed to shine at
 significantly super-Eddington luminosities for a few days after the
 eruption, and when observed again on day 100, it was 2/3 of
 Eddington. Thus, there is no clear indication for the duration of the
 super-Eddington episode, only a lower limit.  Another recent example
 is Nova LMC 1991, which was observed to be super-Eddington for two
 weeks (\cite{LMC91paper}), reaching a truly impressive luminosity as
 high as $2.6 \pm 0.3 \times 10^{39}$ erg s$^{-1}$! As we shall soon
 see, both V1500 Cygni and LMC 1991 were classified as very fast novae
 and hence are marginally useful candidates for a {\em steady state}
 analysis. Another nova, V1974 Cygni 1992, had very beautiful
 bolometric observations (\cite{Shore94}) which could have been
 potentially very useful for the analysis done here. Unfortunately,
 the distance uncertainty of 1.8 to 3.2 kpc (\cite{Paresce95})
 corresponds to more than a factor of 3 uncertainly in the bolometric
 luminosity, which is too large.

\subsubsection{\NLMC}

 \NLMC~occurred, as the name suggests, in the LMC. Thus, its distance
 is known more accurately than the distance of galactic novae. This
 distance, coupled to an extensive multi-wavelength campaign by
 Schwarz et al.~(\jcite{Schwarz98}), resulted with their rather
 accurate finding that \NLMC{ }had an average bolometric luminosity of
 $(3.0\pm0.3)\times 10^{38} \mr{~erg~s^{-1}}$ during the first 45 days
 after visual maximum. By considering that the maximum mass a WD can
 have is $1.4~M_{{\sun}}$, Schwarz et al. (\jcite{Schwarz98})
 concluded that the nova had to be super-Eddington for the long
 duration. We use their data and results for the analysis. This
 includes the evolution of the bolometric luminosity and the effective
 temperature (as opposed to a colour temperature).

\subsubsection{\NFH}

 \NFH~is less clear than \NLMC. Specifically, an accurate distance
 determination was obtained only after using the HST measurement of
 the expanding ejecta (\cite{Gill00}). The reddening which is
 inaccurately known (and is probably in the range
 $E(B-V)=0.64\pm0.16$, \cite{Valle97}) then remains as the main source
 of error in the bolometric luminosity determination. If we take the
 determination of the possible range for the bolometric luminosity by
 Duerbeck (\jcite{Duerbeck92}) and della Valle et
 al.~(\jcite{Valle97}) and correct for the somewhat larger distance
 determined by {Gill} \& {O'Brien} (\jcite{Gill00}), we find that the
 bolometric luminosity on day 6.4, when the first UV measurement was
 taken, is $M_{\mr{bol}}=-7\fm5\pm0\fm2$, namely, $L$ must have been
 more than $2.5\times 10^{38} \mr{~erg~s^{-1}}$.

 Furthermore, since the total mass ejected has a lower bound of
 $5\times 10^{-5}~M_{{\sun}}$, and is probably more of the order of $2
 \times 10^{-4}~M_{{\sun}}$, we should take the kinetic energy of the
 outflow into account. Since the ejecta is moving at about
 $500~\mr{km~s^{-1}}$ (\cite{Gill00}), the ejecta has a minimum
 kinetic energy of $2.5 \times 10^{44}~\mr{erg}$. A lower estimate to
 the contribution of the kinetic energy to the total flux at the base
 of the wind would be to divide the kinetic energy uniformly over the
 duration of the bolometric plateau that lasted about 45 days. This
 gives a minimum contribution of $L_{\mr{kin,min}} = 6.5\times
 10^{37}~\mr{erg~s^{-1}}$. Since the total mass loss is probably
 higher than $5\times 10^{-5}~M_{{\sun}}$ and since the mass loss rate
 was likely to be higher than average when the luminosity was higher
 than the average luminosity, it is likely that $L_{\mr{kin}} >
 L_{\mr{kin,min}}$ on day 6.4.  When we add the kinetic energy to the
 observed bolometric luminosity at day 6.4, we find that
 $L_{\mr{tot}}$ at the base of the wind must have been at least $3.1
 \times 10^{38}\mr{~erg~s^{-1}}$. Clearly, even if the mass of the WD
 is large (which is inconsistent with the eruption not being a very
 fast nova), the luminosities are super-Eddington. The fact that the
 kinetic energy necessarily implies that the nova was super-Eddington
 (at the base of the wind) was already pointed out by Friedjung
 (\jcite{Friedjung87}) who did not have observations on the mass
 ejected but instead used photospheric constraints on $\Md$. Here, we
 have reached the same conclusion without resorting to the
 photospheric analysis. This could be done due to the better distance
 measurements (which are within the error but on the high side of
 previous estimates), and better reddening measurements of della Valle
 et al.~(\jcite{Valle97}).

 We use the data of Friedjung (\jcite{Friedjung87}, also in
 \cite{Friedjung89}) for the behaviour of the temperature, luminosity
 and velocity at infinity. These data are based on the UV measurements
 by Gallagher \& Code (\jcite{Gallagher74}) and in addition include
 the integrated flux that falls outside the UV range observed by
 Gallagher \& Code assuming a black body distribution\footnote{Note
 that when taking this extra flux outside the observed UV range, the
 slow increase followed by a decrease in bolometric luminosity, as
 described by Gallagher \& Code (\jcite{Gallagher74}), turn into a
 slow decrease of the bolometric luminosity.}. We correct the
 luminosities to include the better distance determination by Gill \&
 O'brian (\jcite{Gill00}).

\subsection{Is a steady state model appropriate for the `bolometric
 plateau' of novae?}

 A steady state model is probably appropriate for two reasons:
\begin{enumerate}
 \item Circumstantial evidence: {Kato} \& {Hachisu} (\jcite{Kato94})
 compared their steady state model for winds to the results of a
 dynamical evolution of Prialnik (\jcite{Prialnik86}) and found
 relatively good agreement, implying that a steady state solution is
 valid for most of the nova evolution.

 \item Physical reasons: A wind with a terminal velocity of
 $v_{\infty}$ originates from a typical radius $r_{\st}$ that has an
 escape velocity of order $v_{\infty}$. Two conditions should
 therefore be satisfied for the wind to be in a steady state. First,
 the sub-sonic region beneath the base of the wind should be
 acoustically connected on time scales shorter than the typical
 evolution time of the system. Namely:
\be
t_{\mr{sub-sonic}} \approx {r_{\st}\over v_{s}} \ll t_{\mr{evolution}}.
\ee

The second criterion is that the time it takes the accelerated 
material to reach the photosphere is shorter than the time at 
question:
\be
t_{\mr{super-sonic}} \approx {r_{\mr{ph}}\over v_{\infty}} \ll 
t_{\mr{evolution}},
\ee
 where $t_{\mr{evolution}}$ is the typical time scale for changes in
 the system.  Both \NFH~and \NLMC~satisfy both requirements from days
 4 and 2 onward respectively. Thus, a steady state solution for the
 objects ought to be found for $t_{\mr{evolution}} \approx {\mr
 month}$. As previously mentioned, Novae V1500 Cygni and LMC 1991 do
 not satisfy the required conditions -- they are very fast novae that
 evolve dramatically on a time scale of days, so a steady state wind
 cannot be assumed in their analysis.
\end{enumerate}

\subsection{Application to \NFH}
\label{sec:applNFH}

 Using the described method of analysis, we proceed to fit the data of
 Friedjung (\jcite{Friedjung87}) which includes $T_{\mr{colour}}$,
 $L_\infty$ and $v_\infty$. Given a wind parameter $\func$, a white
 dwarf mass $M_{\mr{WD}}$ and a hydrogen mass fraction $X$, we solve
 for, at each observed point, the values of $\Gamma_0$ and $\Wt$ which
 result with the observed $v_\infty$ and $\Gamma_\infty$. This is done
 iteratively. That is, $\Gamma_0$ and $\Wt$ are assumed to be known
 such that $\Gamma_\infty$ and $v_\infty$ can be calculated according
 to the procedure described at the end of \S\ref{sec:load}. Then,
 $\Gamma_0$ and $\Wt$ are varied until the calculated $\Gamma_\infty$
 and $v_\infty$ agree with observations.  (A function of $\Gamma_0$
 and $\Wt$ is defined such that its value is the error between the
 observations and the obtained values of $\Gamma_\infty$ and
 $v_\infty$. The function is then minimized using a maximum
 `downhill' gradient method).

 The resulting set of $\Gamma_0$ and $\Wt$ for each observation point is then
 used to predict a colour temperature. The predicted temperatures are
 then compared with the observed temperatures (and their error) to
 give a $\chi^2$ number.

\begin{figure} 
\centerline{\epsfig{file=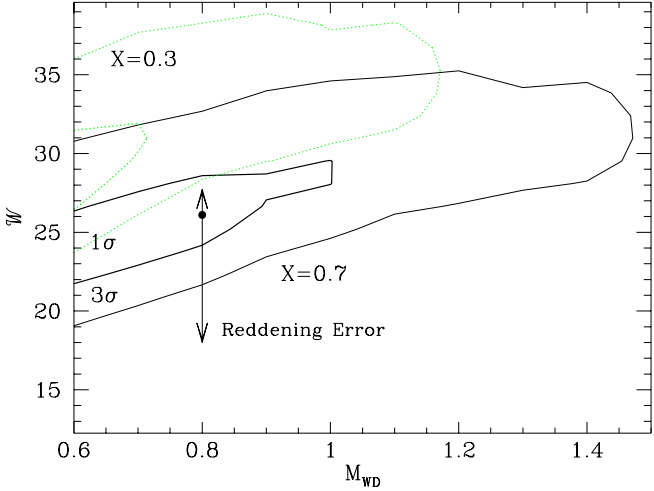,angle=-90,width=\figurewidth}}
\caption{
  The $\chi^2$ fit of the predicted temperature (from the luminosity
  and wind theory) to the observed temperature variations of \NFH, for
  two values of the Hydrogen mass fraction $X$. The 1 and 3 $\sigma$
  statistical variation contours are marked in the $\func$ --
  $M_{\mr{\mr{WD}}}$ plane. Good fits are obtained for reasonable
  parameters. A $\chi^2$ of 4.2 is obtained for $8-2=6$ degrees of
  freedom. The systematic errors are large and arise from the
  inaccurate knowledge of $M_{\mr{\mr{WD}}}$, $X$ and the reddening to
  \NFH. The limit on the systematic variation of the latter is given
  by the arrows.  }
\label{fig:FHchisqr}
\end{figure}

  The results for $X=0.3,0.7$ are depicted in
  Fig.~\ref{fig:FHchisqr}. We find that the possible $1-\sigma$ range
  for $\func$ is about 21.5 to 34. If we allow larger statistical
  variation of $3-\sigma$ then the value of $\func$ can range from 20
  to about 39.5 (if we restrict ourselves to a reasonable WD mass
  range between 0.6 and 1.0 solar masses, as the decay rate $t_3$
  would suggest it is).

 An additional systematic error arises from the inaccuracy of the
 derived reddening of \NFH. This uncertainty is portrayed by the error
 which could increase or decrease the derived $\func$ by roughly +1.8
 or -7.2 respectively.

\begin{figure} 
\centerline{\epsfig{file=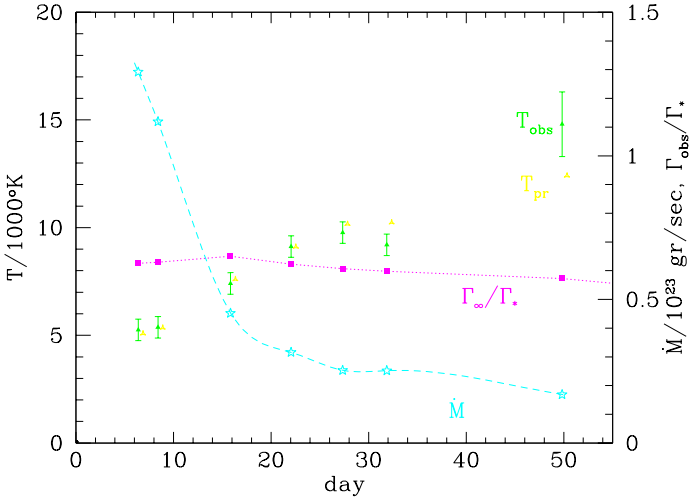,angle=-90,width=\figurewidth}}
\caption{
  The observed colour temperature behaviour $T_{\mr{obs}}$ of
  \NFH~(filled triangles, taken from Friedjung 1987) as compared with
  the predicted colour temperature behaviour $T_{\mr{pr}}$ from the
  observed luminosity and wind model (open triangles, slightly
  offsetted to the right), for the best fit model assuming
  $M_{\mr{\mr{WD}}}=0.8 M_{\sun}$ and $X=0.70$. The additional plots
  are of the mass loss (the stars and dashed line). The mass loss
  integrates to $8\times 10^{-5}M_{{\sun}}$. Finally we plot as filled
  squares the ratio of $\Gamma_{\infty}$ to $\Gamma_0$. Apparently,
  about 40\% of the original radiative luminosity at the base of the
  wind is used-up under these parameters to accelerate the wind and
  compensate for its potential energy. Clumpiness will reduce this
  fraction.}
\label{fig:FHTemp}
\end{figure}

 Fig.~\ref{fig:FHTemp} depicts the observed colour temperature
 evolution and the predicted colour temperature evolution using the
 best fit values of $\func$, $\Gamma_\infty(t)$ and $\Wt(t)$, and the
 nominal values of $X=0.7$ and $M_{\mr{\mr{WD}}}=0.8 M_{{\sun}}$. Good
 agreement between the predicted temperature using the SEW theory and
 the observed colour temperature can be obtained, clearly
 demonstrating that the theory is consistent.

 The temporal evolution of the observed luminosity, as given in the
 same figure, can be integrated to obtain a total mass loss of $8
 \times 10^{-5} M_{{\sun}}$. The main source of uncertainty in this
 figure is again the reddening which could increase or decrease the
 mass loss by about +10\% and -30\% respectively. This result is
 completely consistent with the observational range of $(2 - 20)
 \times 10^{-5} M_{{\sun}}$ (\cite{Nasa1}, \cite{Gill00}).

 An important point which should be considered is that the wind is
 likely to be clumpy, a fact which the analysis thus far did not take
 into account. In the region around the sonic point, the optical depth
 of typical clumps is expected to be of order unity. Above this point,
 typical clumps are therefore expected to be optically thin. Thus, the
 main effect of the clumpiness is to increase the effective absorptive
 opacity (since one will then obtain that $\kappa_{\mr ab} \propto
 \sqrt{\left< \rho^2 \right>} >\sqrt{\left< \rho \right>^2}$). The
 effective opacity which is proportional to the geometric mean of the
 scattering and absorptive opacities will increase as $\sqrt{\clump}$
 where $\clump$ is a clumpiness parameter:
\begin{equation}
\clump \equiv \sqrt{\left< \rho^2 \right> / \left< \rho \right>^2}.
\end{equation}
 Moreover, from Bath \& Shaviv (\jcite{Bath76}) we have that for a
 fixed temperature, outflow velocity and luminosity, the inferred mass
 loss rate is proportional to $\kappa^{-3/2}$. Thus, a first guess
 would be that by assuming homogeneity we over estimate $\func$ by
 roughly a factor of $\clump^{3/4}$.  A more detailed analysis which
 repeats the fitting actually shows that the power law relation is
 over estimated by about 10-15\%. That is to say, $\func$ is over
 estimated by about $\clump^{0.67}$ for $M_{\mr WD} \sim (0.8 - 1.0)
 M_{\sun}$.

 We do not know what the value of $\clump$ should be. A rough
 guesstimate would be to take the value found in another type of
 system in which strong optically thick winds are observed, namely, WR
 stars. In binary systems in which one of the stars is a WR, several
 measurements of the mass loss in the winds can be performed. Some
 (which are usually the more difficult ones) measure the actual mass
 loss $\Md$ (for example, using scattering induced polarizations or
 measurement of the period slowdown of the binary, $\dot{P}$) while
 others measure $\Md \clump$ (for example, using radio measurements of
 the free-free emission). In the case of V444 Cygni, a value of
 $\clump \approx 3$ is obtained (\cite{LML93}). If we adopt this value
 as a typical one, then the inferred range of $\clump^{3/4}\func$,
 which is 12.5 - 41.5, corresponds to $\func =5.5 - 18 $.

\subsection{Application to \NLMC}
 
 \NLMC~has better data than \NFH~in several respects. First, the
 absolute luminosity is known with higher accuracy. Second, the
 temperature obtained by Schwarz et al.~(\jcite{Schwarz98}) is an
 effective temperature and not the colour temperature. Consequently,
 the analysis does not depend on the absorptive opacity and therefore
 the clumpiness of the wind. The main draw back however, is that
 \NLMC~does not have adequate measurements of the evolution of the
 velocity of the outflow. The velocity of $1800 \pm
 200~\rm{km~s^{-1}}$ adopted by Schwarz et al.~(\jcite{Schwarz98}) is
 based on the measurements of optical emission lines that are also
 consistent with the subsequent observations of `Orion' emission lines
 when the nova was optically thin. There are no available records to
 velocities derived from the principal (absorptive) spectrum.

 An important question should be raised. Is the velocity of the the
 photospheric material given by the velocity of the diffuse enhanced
 spectrum or by the principal spectrum? The answer is probably the
 latter. Irrespective of the theoretical argumentation for which
 spectrum is formed closer to the photosphere, the circumstantial
 evidence points to the principal spectrum. This is because the
 velocities obtained from the principal spectrum are similar to those
 obtained by measurements of the nebular expansion years after the
 eruption. Namely, the principal spectrum provides the velocity of the
 bulk of the material while the diffuse enhanced spectrum probably
 gives the velocity of the small amount of material that is initially
 injected at high speeds. In \NFH~for example, the principal spectrum
 gives velocities in the range $670$ to $770~\mr{km~s^{-1}}$. The
 diffuse enhanced spectrum is in the range $1300$ to
 $1900~\mr{km~s^{-1}}$. Clearly, the principal spectrum is closer to
 the observed expansion velocity of the nebula at
 $490\pm20\mr{~km~s^{-1}}$ (\cite{Gill00}).

 In order to obtain an estimate for what the principal velocity is, we
 take the two relations that relate $t_3$ to $v_{\mr{principal}}$ and
 $v_{\mr{diffuse}}$ (e.g., \cite{Warner89}) and relate the two
 velocities. The $1800 \pm 200~\mr{km~s^{-1}}$ observed for the
 emission lines, then corresponds to a principal spectrum of $900\pm
 150 \mr{~km~s^{-1}}$.

 Taking the above considerations into account, the analysis proceeds
 in a manner similar to that of \NFH~with a few notable exceptions.

 In addition to $\func$, $M_{\mr{WD}}$ and $X$, we leave the velocity
 $v_\infty$ as a free parameter and let it vary from 750 km s$^{-1}$
 to 2000 km s$^{-1}$, corresponding to the complete range of possible
 velocities, without actually any prejudice for or against which
 spectral velocities should be taken. We assume the velocity to be
 constant with time, which towards the end of the eruption, when small
 mass outflows are expected (and therefore a smaller observational
 footprint), can be a bad approximation. We therefore fit the data
 only to the first month of observations.

\begin{figure} 
\centerline{\epsfig{file=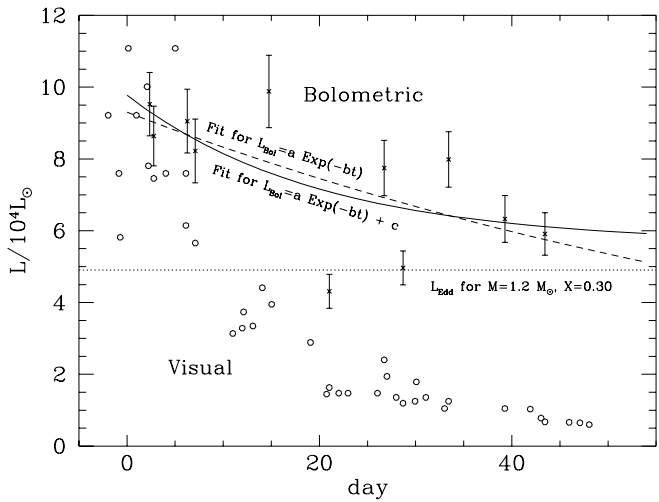,angle=-90,width=\figurewidth}}
\caption{
  The visual and bolometric luminosity of \NLMC~taken from Schwarz et
  al.~(1998). Open circles are the visual luminosities. The measured 
  bolometric luminosities are marked by {\sf x}'s and an error bar.
  Because of the moderate decay rate of the nova, its
  mass should be of order or less than about $1.2M_\odot$. $X=0.3$ is
  the typical hydrogen mass fraction in nova ejecta.  The dashed line is a
  least squares fit to an exponential decay for the bolometric
  luminosity, giving $L_{\mr{bol}}=9.3\times 10^4 L_\odot\exp (-t/90.5
  {\mr~days})$. The solid line is the best fit for a luminosity
  behaviour for an exponential decay plus a constant, giving:
  $L_{\mr{bo}l}=4.13\times 10^4 L_\odot\exp (-t/20.0~{\mr
  days})+5.64\times 10^4 L_\odot$. Evidently, the nova was
  super-Eddington for a long duration. Unfortunately, adequate
  measurements of the evolution of the velocity as a function of time
  do not exist. }
\label{fig:LMC88Bol}
\end{figure}

 Since the dimensionality of the free parameter space is too large, we
 do not plot the contours of $\chi^2$ but instead find and plot the
 value of $\func$ which minimizes it given $M_{\mr{WD}}$, $X$ and
 $v_\infty$.  Moreover, since the observed luminosity exhibits rather
 large variations, as is apparent in Fig.~\ref{fig:LMC88Bol}, a
 smoothed functional behaviour is adopted. A fit of the form
 $L_{\mr{bol}} = a \exp\left(-b t\right) +c$ yields the best fit.

\begin{figure} 
\centerline{\epsfig{file=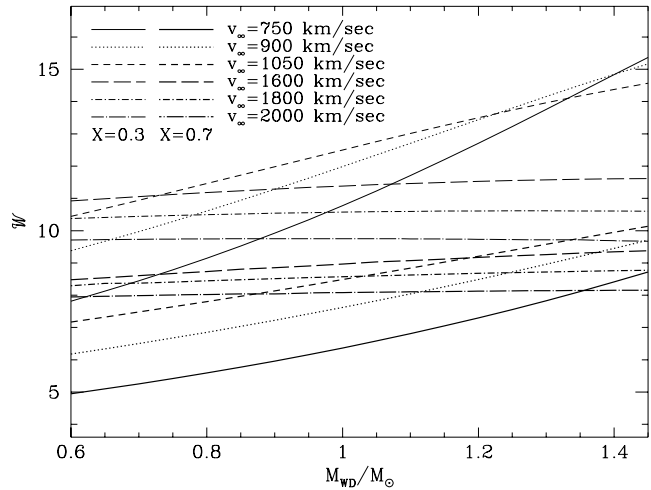,angle=-90,width=\figurewidth}}
 \caption{ The best fit values of $\func$ for \NLMC~found for
 different values of the H mass fraction $X$ and different assumed
 constant velocity for the outflow, as a function of the assumed WD
 mass. Values ranging between 5 and 15 are possible. }
\label{fig:LMCW}
\end{figure}

 The result of the aforementioned procedure are depicted in
 Fig.~\ref{fig:LMCW}.  Without any prejudice for $M_{\mr{\mr{WD}}}$,
 $X$ or $v_{\infty}$ (except for limiting them to a reasonable range
 of $0.6 M_{\sun}<M_{\mr{\mr{WD}}}<1.4 M_{\sun}$, $750~{\mr
 km~s}^{-1}<v_\infty< 2000~{\mr km~s}^{-1}$ and $0.3<X<0.7$), one
 finds that the wind model can adequately explain the results with a
 wind parameter in the range:
\be
\func_{\mr{LMC}} = 10.0 \pm 5 .
\ee
This result is consistent with both the one obtained from \NFH~and the
one expected from the SEW theory for $\beta$ somewhat smaller than
unity and/or $\alpha$ either closer to 1 or much closer to $0$ than
the nominal value of $\alpha=1/2$.

\subsection{The great eruption of $\eta$ Carinae}

 The case of $\eta$ Carinae is quite different from the novae
 analyzed. First, the systems are entirely different. Instead of a
 solar mass type WD, $\eta$ Car is a blue super-giant with a mass of
 order of $60$ to $100M_{\sun}$. While the super-Eddington phase of
 novae lasts of order a few months during which there is mass loss
 with a typical rate of $10^{-3}~M_{\sun}~\mr{yr}^{-1}$, the giant
 eruption of $\eta$ Car which started around 1840, lasted for 20 years
 and exhibited a mass loss rate of order
 $10^{-1}~M_{\sun}~\mr{yr}^{-1}$.

 A second difference appears in the way the analysis proceeds. While
 the novae analyzed had a detailed evolution of the temperature and
 luminosity, all that we have for $\eta$ Car is an estimated average
 luminosity during the 20 year eruption and an estimate for the
 integrated mass loss.  Yet we apply the theory and stress that the
 very same theory is applicable to a wide range of systems.

 Following Davidson (\jcite{Davidson99}), we adopt the following
 parameters for the eruption of $\eta$ Car: Duration of 20
 years. Integrated radiated flux of
 $10^{49.3\pm0.3}~\mr{erg}$. Ejected mass of $1$ to $3M_{{\sun}}$, and
 a mass velocity at infinity of $650~\mr{km~s^{-1}}$.  The mass
 adopted is $80 \pm 20 M_{\sun}$ which can correspond to the estimate
 of $\sim 60 M_{\sun}$ if it is a part of a double star and $\sim 100
 M_{\sun}$ if it is single.

 To obtain $\func$ from the observables, we first assume that the mass
 loss rate was constant and equal to $\Md=(1\pm0.5)M_{\sun}~{\mr
 yr}^{-1}$. Given the aforementioned nominal values for the
 luminosity, velocity at infinity, stellar mass and mass loss rate, we
 can iterate for the values of $\Gamma_0$ and $\Wt$ according to the
 procedure outlined at the end of \S\ref{sec:load} and the fitting
 done to each point of \NFH~ in the beginning of
 \S\ref{sec:applNFH}. The difference between the cases of $\eta$ Car
 and \NFH~is that here we only have one data point (corresponding to
 the `averaged' eruption) and that it does not include a temperature
 measurement. Therefore, we do not have enough observables to obtain a
 $\chi^2$ fit. Namely, we have the correct number free of parameters
 to fit the data exactly. The set of nominal values for the
 observables therefore results with a particular value for $\func$. To
 obtain the error in the estimate, we take each observable and change
 it to the extremes allowed by its error bar. This gives the error in
 $\func$ induced by that particular observable. The errors from the
 different observables are then added in quadrature.

 The wind parameter that we find is
\be
 \func = 16.2 \pm 12.
\ee
 The greatest contribution to the error arises from the inaccuracy of
 the average luminosity while the second largest contribution, which
 is half the size of the first, comes from the inaccuracy of the mass
 ejected.

 Although the errors in $\func$ are larger than those obtained for
 \NLMC, the range obtained from $\eta$ Car includes entirely the range
 obtained from \NLMC. Namely, the results are consistent and the value
 obtained from \NLMC~should be taken as the best estimate for $\func$. 

\subsection{Possibility of Super-Eddington fluxes}

 Although we cannot say at the moment anything quantitative about the
 steady state luminosity that will be attained without proper
 knowledge of the properties of the porous atmosphere, we can
 understand why super-Eddington fluxes are a natural result.

 The steady state luminosity at the post-maximum of novae is generally
 supposed to be given or at least approximated by the core-mass
 luminosity relation (CMLR, e.g., \cite{Tuchman98}) which describes
 systems in which a burning shell is situated on an inert (and hence
 fixed) core (\cite{Paczynski70}). The CMLR however, saturates at
 the Eddington luminosity. So, how can it describe cases which were
 clearly super-Eddington?

 A clear physical understanding of the CMLR can be found in {Tuchman}
 {et~al.} (\jcite{Tuchman83}) who studied a system composed of an
 inert core on top of which there is a burning shell, a radiative
 layer with sharp gradients and a convective layer on top of
 that. They showed that the conditions above the radiative layer are
 unimportant for the determination of the burning luminosity (thus,
 the mass of the envelope is unimportant). Only the burning layer and
 the radiative layer on top of it affect the luminosity. How would a
 porous atmosphere change that? Since any inhomogeneities are expected
 to form in the top part of the atmosphere (above the convective
 layer), it does not change the analysis of {Tuchman} {et~al.}
 (\jcite{Tuchman83}). That is the case only as long as a consistent
 solution for the top part of the atmosphere is obtainable.

 However, as the mass of the WD increases and with it the Eddington
 parameter, the top part of the atmosphere becomes unstable against
 formation of inhomogeneities and once these are formed, allows a
 larger radiative flux for the same average temperature and density
 gradients.  If this larger flux is close enough (or larger than) the
 Eddington luminosity, one of the main assumptions in the analysis of
 {Tuchman} {et~al.} (\jcite{Tuchman83}) breaks down because the
 radiative layer above the burning shell necessarily has to become
 convective (\cite{JSO73}).  Thus, an additional branch for the
 mass-core luminosity relation becomes possible in which the top part
 of the atmosphere is inhomogeneous and below it, all the way down to
 the burning shell, the envelope is convective.  Since the luminosity
 now depends on the photospheric conditions (the convective layer
 adjusts itself to `relate' the conditions below and on top of it),
 the core-mass luminosity relation can lose its insensitivity to the
 parameters of the envelope.

\subsection{The `Transition Phase'}

 One of the seemingly odd behaviour displayed by a large majority of
 the classical nova eruptions is a transition phase. If it appears, it
 starts once the visual magnitude has decayed by 3 to 4
 magnitudes. During the transition phase, the light curve can display
 strong deepening, quasi-periodic oscillations, erratic changes or
 other complicated behaviour.

 The origin of the transition phase is not clear and more than one
 explanation had been suggested. For example, the transition phase
 roughly corresponds to the stage when the photosphere has shrunk to
 the size of the binary separation such that the companion star can
 stir up the envelope to produce non trivial behaviour.

 The SEW theory naturally introduces another explanation for the
 transition phase. If we look at the $\Gamma_0$ - $\Wt$ trajectory of
 \NFH~in Fig.~\ref{fig:contours}, one cannot avoid the extrapolation
 of the trajectory into the zone of `no steady state
 configuration'. What is this region?

 The mass loss rate is determined by the luminosity. The mechanism
 that generates this wind does not depend on whether the luminosity is
 sufficient or not to carry the material from the sonic point to
 infinity. If the radius of the sonic point is too small and the
 escape velocity too high, the wind simply stagnates before leaving
 the potential well. As predicted by Owocki \& Gayley
 (\jcite{Owocki97}), no steady state solution for the wind exists
 under such conditions. In other words, the wind model predicts that
 no steady state could be reached if the radius falls too
 quickly. This appears to be a likely scenario in the case of \NFH~if
 one extrapolates its trajectory seen in Fig.~\ref{fig:contours},
 and indeed, \NFH~was observed to have a transition phase in which the
 luminosity faded dramatically.

 Both the fading and the erratic or quasi-periodic behaviour seem like
 a natural result here. From Fig.~\ref{fig:contours}, it is apparent
 that $\Gamma_\infty$ can fade dramatically before entering the
 transition phase. This arises from the fact that close to the
 transition phase almost all of the flux is used to pay for the
 potential toll. Once in the `domain of no steady state' the non
 trivial 2D or 3D flows that must result could potentially result with
 non trivial variability.

 The faded and/or variable luminosity phase is expected to end when
 the luminosity at the base of the wind falls below the Eddington
 limit, shutting off the SEW, at which point the `naked' white dwarf
 should emerge.

\subsection{General Mass Loss of Novae}
 We have previously treated two specific novae for which the temporal
 evolution of the luminosity and temperature is known in detail. We
 now turn to describe the nova population in general for which only
 the average behaviour is known. To do so, we create a template nova
 as a function of white dwarf mass and explore its properties. Its
 predicted mass loss is then compared with the observed integrated
 mass loss. Clearly, we expect the theoretical prediction to provide
 the guide line to which the {\em average} observed behaviour should
 be compared to.

 Let us first estimate the average mass loss in novae. To this goal,
 we use the average trends in the functional behaviour of the
 luminosity, decay time and velocity as a function of white dwarf mass
 to predict the integrated mass loss during the super-Eddington
 episode.  Specifically, we use the following:
\begin{enumerate}
 \item We use the average relation between $t_3$, the time it takes
 for the visual luminosity to decay by 3 magnitudes, to the mass of
 the WD, as is given by Livio (\jcite{Livio92}).

 \item We assume that the photospheric velocity is given by the
 velocity of the principal spectrum. This velocity has an average
 relation to the decay time $t_3$, as given for example by Warner
 (\jcite{Warner89}).

 \item We take the peak bolometric luminosity to be the peak visual
 luminosity. This is permissible since novae's maxima are generally in
 the visual. $M_v$ is related to $t_3$, and therefore to
 $M_{\mr{\mr{WD}}}$, by relations given by Livio
 (\jcite{Livio92}). Using $M_{\mr{\mr{WD}}}$, $v_\infty$ and
 $\Gamma_{\infty}$, we find the Eddington parameter $\Gamma_0$ at the
 base of the wind during the peak brightness. We do so using the
 relations given in section \ref{sec:load}.

 \item We assume that the bolometric luminosity at the {\em base}
 decays exponentially (or the magnitude linearly). That is, it has the
 form: $L (t) = \Gamma_0 \L_{\mr{Edd}} \exp(-t/\tau)$. Since we expect
 the transition phase to arise when the luminosity approaches the
 Eddington luminosity, and since the transition phase usually sets in
 after the visual decayed by 3 to 4 magnitudes, or about 3.5 on
 average, we can relate $t_3$ to the exponential decay constant of the
 base luminosity $\tau$:
\be
  \tau \approx {(3.5/3) t_3 \over \ln (\Gamma_0)} .
\ee
 Few words of explanation are in order. It is conceivable that as the
 nova eruption proceeds, since the amount of fuel is fixed and all of
 it burns simultaneously, we do not get a strict steady state
 cigar-type burning and fixed bolometric luminosity.  The bolometric
 luminosity must change gradually with time. It is customary to assume
 as a first approximation that $L_{\mr{bol}}={\mr constant} $
 (\cite{Bath76}). However, a more realistic treatment that more
 accurately describes the observations, is to assume a gradual, though
 slow, decline expressed as an exponential decay with a long time
 scale (see for instance Fig.~\ref{fig:LMC88Bol}). For comparison, we
 also repeat the whole calculation using a linear decline of the form
 $L(t) = \Gamma_0 \L_{\mr{Edd}} (1-t/\tau)$ and show that the exact
 form of decay is not critical.

\item If we integrate the mass loss rate given by
the wind theory (eq.~\ref{eq:masslossrate}), we find that the total
ejected material during the super-Eddington episode is:
\ba
 M_{\mr{ejecta}} &=& \int_0^{t_E} \Md dt = {\func \over c v_s}
 \int_0^{t_E} \left(L - \L_{\mr{Edd}} \right) dt \nonumber \\ &
 \approx & {\func \over c v_{s}}\L_{\mr{Edd}} \tau \left( \Gamma_0 -1
 -\ln(\Gamma_0) \right),
\ea
where $t_{E}$ is the time it takes the bolometric luminosity to
decline to the Eddington limit.
\end{enumerate}

\begin{figure} 
\centerline{\epsfig{file=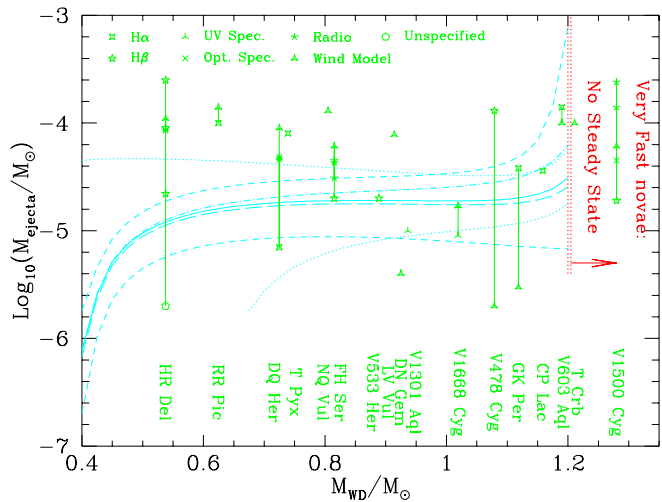,angle=-90,width=\figurewidth}}
 \caption{ Nova ejected shell mass vs.~mass of WD. The symbols
 describe the measured mass loss for different observed novae as
 compiled by Hack et al.~(1993).  Large measurement errors are
 apparent in cases where more than one measurement exists (in which
 case a vertical line connects the points).  One should also note that
 some measurements overestimate the mass loss when the wind is clumpy,
 as it is predicted to be.  The solid line describes the mass loss
 obtained by the wind model assuming a nominal value of $\func=10.0$
 as obtained from \NLMC, following the procedure given in the text.
 The long dashed line is obtained when changing the Hydrogen mass
 fraction from 0.7 to 0.3.  The upper (and lower) short dashed lines
 are obtained when taking a value for $\func$ which is higher (lower)
 by 5.0, while the dash-dotted line arises when the assumed
 exponential decay is replaced by a linear decay.  The dotted lines
 arise when taking into account the natural scatter in the $t_3$ --
 $M_{\mr{\mr{WD}}}$ relation obtained by Livio (1992).  }
\label{fig:masslossmass}
\end{figure}

 The results are plotted in Fig.~\ref{fig:masslossmass} as a function
 of the mass of the white dwarf, together with the observed
 determinations of ejecta masses, as was complied by {Hack} {\em
 et~al.}~(\jcite{Nasa1}). We first notice that in those cases where
 multiple determinations were obtained, there are large differences
 between measurements. These differences should therefore be taken as
 the typical `error' in cases where only one determination was
 performed. The different theoretical predictions correspond to
 changing the wind parameter $\func$ to within the possible range of
 $10.0 \pm 5$, when assuming a linear instead of an exponential decay,
 and when taking into account the natural scatter in the $t_3$ --
 $M_{\mr{\mr{WD}}}$ relation of Livio (\jcite{Livio92}).

 To within the large uncertainties in the ejecta mass determinations
 (which have a $1-\sigma$ scatter of 0.54 dex), the prediction and
 observations appear to be in good agreement.  The super-Eddington
 episode of novae could account for the bulk, if not all, of the
 ejected material as a function of $M_{\mr{WD}}$.  One should
 nevertheless note that the logarithmic average of the observations:
 $\left< \log_{10} (M_{\mr{ejecta}} / M_{\sun}~{\mr{yr^{-1}}})\right>
 = -4.40 \pm 0.09$ is on the upper side but still within the allowed
 region predicted by the SEW theory (which has only a small functional
 dependence on $M_{\mr{\mr{WD}}}$). Since some of the ejecta mass
 determinations are susceptible to clumpiness, taking the latter into
 account should tend to reduce the logarithmic average. If $\clump
 \approx 3$ is a typical value and if characteristically about a half
 of the measurements actually measure $M_{\mr{ejecta}}\clump$, then the
 average of the mass loss would be smaller by about a 0.25 dex.

 Several additional conclusions can be drawn from
 Fig.~\ref{fig:masslossmass}.  The plateau in the predicted mass loss,
 where the mass loss does not depend on the mass of the white dwarf,
 extends from $M_{\mr{WD}}=0.5M_{\odot}$ to $1.2M_{\odot}$. This
 appears to agree with observations.  Moreover, the extrapolation to
 WD masses beyond $1.2M_{\odot}$ agrees with the mass loss observed
 from the very fast nova V1500 Cygni, however, this extrapolation
 should be done cautiously because the assumption of a steady state
 wind is not strictly valid.

 The plateau predicted by the SEW and seen in the observations is
 counter to theoretical predictions of the TNR process in which the
 general trend should be a smaller mass loss with larger WD
 masses. This theoretical trend arises from the fact that more massive
 WDs are more compact and ignite the TNR after significantly less
 material is accreted. For some reason, this trend is not seen.  The
 current TNR theory of novae tends to predict values which are about
 half an order to an order of magnitude smaller than the typical
 observations for large WD masses (e.g., \cite{Starrfield99},
 \cite{Prialnik95}, \cite{Starrfield98}). The implementation of the
 SEW theory in numerical simulations of thermonuclear runaways in now
 underway with the purpose of finding the luminosity at the base of
 the wind self consistently as well as to see to what extent the
 incorporation of the SEW changes the predicted amount of ejected
 material.
 
\subsection{A `Constant Bolometric Flux'}
 Let us return to Fig.~\ref{fig:contours}.  Recall that lines of
 constant radius are close to being vertical.  One can see that for
 moderately `loaded' winds and $\Gamma_0$ of a few, an evolution in
 which the temperature increases but the apparent luminosity remains
 constant is possible if the radius does not decrease dramatically.
 Under such conditions, the luminosity at the base of the wind does
 fall off to get a higher and higher temperature.  However, the lower
 mass loss predicted implies that less energy is needed to accelerate
 the material to infinity and so a larger fraction of the base
 luminosity remains after the wind has been accelerated.

 This could explain for example the constant bolometric luminosity
 observed for V1974 Cygni 1992 (\cite{Shore94}) and supports the early
 working hypothesis (\cite{Bath76}, \cite{Gallagher76}) that novae
 evolve with a bolometric luminosity which is almost constant, or at
 least one that does not vary dramatically.

 Interestingly, depending on the load of the wind, the constant
 apparent bolometric flux can be either super or sub-Eddington. That
 is, even if an apparent sub-Eddington luminosity is observed, the
 system could still have been super-Eddington over dynamically long
 durations. In such cases, the kinetic and potential energies of the
 flow are important players in the total energy budget.

\subsection{Clumpiness of the wind}

 Clumpiness was already mentioned on several occasions. Since it is
 important, it deserves a more concentrated discussion.

\begin{figure} 
\centerline{\epsfig{file=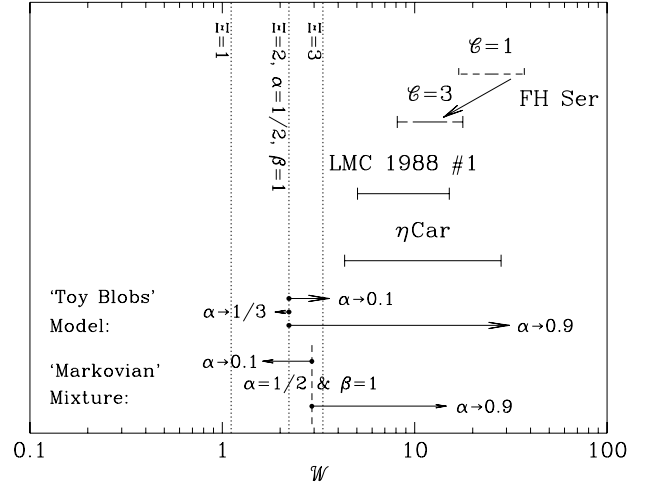,angle=-90,width=\figurewidth}}
 \caption{Comparison of the wind parameter obtained from three
 independent objects. In the case of FH Ser, the result
 $\clump^{0.67}\func$ constitutes an upper limit since any clumpiness
 in the wind will mimic a larger inferred value for $\func$ if
 homogeneity (i.e., $\clump=1$) is assumed. The value of $\clump=3$ is
 a guesstimate from the clumpiness observed in WR winds.  Since the
 wind is expected to be clumpy, the larger value obtained is a good
 indication that the analysis and the model are consistent. The
 vertical dotted lines describe the theoretical prediction and its
 uncertainties using the simple `blob' model. For nominal values for
 the unknown geometrical parameters ($\alpha=1/2$, $\beta=1$ and
 $\Xi=2$), the central line is obtained using eqs.~(\ref{eq:B}) and
 (\ref{eq:WAB}). The two additional dotted lines represent the
 uncertainties arriving from the geometrical parameter $\Xi$
 describing the surface to volume ratio in units of the size of the
 blobs, which could reasonable vary from $\Xi\sim 1$ to $\Xi\sim
 3$. The arrows represent the change in the predicted $\func$ when
 changing the volume fraction of the dense regions. When large volume
 asymmetry is present, the value of $\func$ increases.  It is
 minimized for $\alpha=1/3$. $\beta$ can shift $\func$ in both
 directions. The dashed vertical line represents the prediction using
 the Markovian results (eq.~\ref{eq:Markovian_res}). Clearly, the
 results from the three objects are consistent with each other. They
 are somewhat larger than the theoretical prediction using nominal
 values ($\beta = 1$, $\alpha=1/2$) but they do agree with the
 predictions using geometrical parameters well within reasonable
 bounds (e.g., {$\alpha \sim 0.8$} or larger). }
\label{fig:W}
\end{figure}

 SEWs are expected to be {\em always} clumpy as they are generated by
 an inhomogeneous atmosphere. The results for $\clump^{0.67}\func $
 for \NFH~as compared to the results for $\func$ of \NLMC~do indicate
 that the winds are clumpy and that the clumpiness factor could be
 similar to that already observed in the optically thick winds of
 Wolf-Rayet stars, namely, $\clump \sim 3$. This can be seen in
 Fig.~\ref{fig:W} which depicts the $\func$ obtained in the three
 discussed systems.

 Clumpiness is also important because it tends to offset the estimates
 for the mass ejected from novae and other objects by overestimating
 it. This arises because many emission processes are more efficient at
 higher densities. For example, the opacity per unit mass of free-free
 emission is proportional to $\rho$. Thus, if the material is clumpy,
 then the higher density material is more efficient in producing the
 observed radiation than the case of the material spread evenly over 
 the entire  volume.

 What is the expected size of the clumps? Since they originate from
 the inhomogeneities at the base of the wind, one needs to know the
 typical size $d$ of the atmospheric perturbations at this point. For
 the instabilities found to operate in Thomson scattering atmosphere
 (\cite{instabilities}), the typical size is of the order of the size
 of the pressure scale height $l_p$ (say, $\beta l_p$). As the wind
 expands from its base, the clumps will keep their angular extent
 relative to the star. Thus, the spherical harmonic $\ell$ at which
 the structure should peek should be of order:
\be
 \ell \approx 2 \pi {R \over d} = 2 \pi {R \over \beta l_p} \approx
 {\pi(1-\Gamma_{\mr{eff},\infty})\over \beta\nu} {v_{\mr{esc}}^2 \over
 v_{s}^2}, \quad \beta \sim {\cal O}(1)
\ee
 where $v_{s}$ is the speed of sound at the base of the wind, which is
 higher than that at the photosphere.  The typical number obtained for
 novae, is $\ell \sim 100$. This is a large number and it implies that
 the ejecta has many small clumps in it.  Since the perturbations are
 expected to be dynamic and change over a sound crossing time, their
 vertical extent should be of order $(v_\infty /v_{s})\beta l_p$, at
 large distances, after the velocity dispersion had time to disperse
 the blobs vertically. The horizontal dispersion arising from
 velocities of order of the speed of sound are not important because
 of the horizontal expansion induced by rarefaction in the spherical
 geometry.

\section{Discussion \& Summary}

 The existence of steady state super-Eddington outflows is an
 observational fact. One therefore needs to explain on one hand how
 atmospheres can sustain a super-Eddington state and on the other, one
 needs to understand the winds that they generate.

 We have tried to present the following coherent picture: Homogeneous
 atmospheres becomes inhomogeneous as the the radiative flux
 approaches the Eddington limit. This is due to a plethora of
 instabilities. The particular governing instability depends on the
 details of the atmosphere. As a consequence of the inhomogeneity, the
 effective opacity is reduced as it is easier for the radiation to
 escape, and consequently, the effective Eddington limit increases.

 Super-Eddington configurations are now possible because the bulk of
 the atmosphere is effectively sub-Eddington so that the average is 
 also sub-Eddington. Very deep layers advect
 the excess total luminosity above Eddington by convection. Higher in
 the atmosphere, where convection is inefficient, the Eddington limit
 is effectively increased due to the reduced effective opacity. The
 top part of the atmosphere, where perturbations of order of the scale
 height become optically thin, has however to remain
 super-Eddington. Thus, these layers are pushed off by a continuum
 driven wind.

 By identifying the location of the critical point of the outflow, one
 can obtain a mass-loss luminosity relation. The relation, given by
 eq.~(\ref{eq:masslossrate}) is the main result of the paper. Adding
 to eq.~(\ref{eq:masslossrate}) the basic results for optically thick
 winds eq.~(\ref{eq:vel-inf}) (\cite{Owocki97}) and
 eq.~(\ref{eq:BSTeff}) (\cite{Bath76}) or eq.~(\ref{eq:BSTcolour})
 (\cite{Bath78}), we are left with one free universal dimensionless
 function $\func$. It was shown that $\func$ depends primarily on the
 geometrical characteristics of the inhomogeneous atmosphere from
 which the wind emerges. Having no detailed information on the
 geometrical properties, we {\em assumed} that they are independent of
 $\Gamma$, so that $\func$ becomes a constant.

 To check these  ideas, we analyze 2 novae as well as the massive star
 $\eta$ Carinae for which sufficiently detailed and accurate 
 data is available. Although the two types of systems are notably
 different, as they have masses, luminosities and mass loss rates which
 differ by orders of magnitude, the wind mass loss and the wind
 parameter are found to be in agreement with the theoretical
 expectation:
\be
 \Md = \func {\L_{\mr{Edd}} \over c v_s} (\Gamma-1),~{\rm with}~\func
 = 10.0 \pm 5.
\ee
 The evolution of the temperature predicted from the luminosity agrees
 well with the temperature measured directly in the two novae.

 Another interesting agreement is the consistency with
 clumping. Clumping in SEWs is a natural prediction since the
 atmospheric layers beneath the sonic point are predicted to be
 inhomogeneous. Moreover, clumpiness is a necessary ingredient in the
 present theory that allows super-Eddington luminosities. The present
 theory predicts therefore, that SEWs are clumpy.  In the analysis of
 \NFH, the wind parameter obtained is coupled to the clumpiness of the
 wind since a clumpy wind with the same observed colour temperature
 will have a lower inferred mass loss. The lower $\Md$ reduces the
 measured wind parameter towards that obtained for \NLMC~when the
 typical clumpiness factors seen in WR winds are taken into account.

 The consistency of all the results is best demonstrated in
 Fig.~\ref{fig:W} which shows that $\func$ obtained from the three
 different objects is consistent with one another. Although $\func$
 obtained is somewhat larger than the theoretical prediction using
 nominal values for the geometrical characteristics, the uncertainties
 in the latter are large enough to comfortably accommodate the range
 of `measured' $\func$'s.

 An additional way of viewing the results is as follows.  The SEW
 theory predicts a range of $\func$ which due to uncertainties can
 span at least an order of magnitude.  The analysis of \NLMC~yields a
 $\func$ which agrees with this range.  The analysis of \NFH~yields a
 $\func$ which too agrees with this range if clumpiness, which is
 predicted, is included.  Next, instead of finding the values of
 $\func$ which best fit $\eta$ Car, we could have asked the opposite
 question.  Using the SEW theory which was `normalized' using \NLMC,
 we could have calculated what the observed mass loss should have been
 according to the SEW theory, and find out that its prediction is
 completely consistent with the observed mass loss, even though the
 object's characteristics are different from novae by several of
 orders of magnitude.

 We also identify the occurrence of the `transition phase' observed in
 more than two thirds of the novae with the advance of the atmosphere
 into the `no steady state region'.  As the nova explosion progresses,
 its luminosity and radius decline. However, if the radius decreases
 too quickly, at some point the predicted SEW will be too heavy for
 the luminosity at the base to push mass to infinity. No steady state
 will then exist. The inconsistency might explain the strange
 behaviours observed in the transition phase of different novae.

 The wind model presented is by no means a complete theory for novae
 since it cannot predict {\em ab initio} the luminosity at the base of
 the wind.  To obtain the latter, one needs to solve for the complete
 evolution of a nova taking into account the fact that the nova's
 atmosphere becomes porous and feels a reduced effective opacity.  One
 expects that the lowered opacity increases the luminosity obtained in
 the core-mass luminosity relation, and super-Eddington values
 therefore arise naturally.  Preliminary numerical solution of the
 stellar structure equations appear to confirm this prediction.

 In order to fully understand the behaviour of the atmosphere that
 drives these winds, a 2D or even 3D numerical simulations on scales
 of the order of few optical depths are unavoidable because of the
 intrinsic nonlinear properties of the problem.  These simulations are
 underway and will allow one to study in detail the physics
 controlling the mass outflow.

\section*{Acknowledgments}

The author would like to thank useful comments from Stan Owocki, Ken
Gayley and CITA for the fellowship supporting him. The author is
particularly grateful to the referee John Castor for the very helpful
suggestions that improved the manuscript.


\appendix
\section{The Effective opacity when not fully in the
 optically thick or thin limits}

 We proceed here to calculate the corrections to the effective opacity
 in systems which are neither fully in the optically thick limit nor
 fully in the optically thin limit.  This analysis is presented to
 strengthen the heuristic arguments given in the section calculating
 the location of the critical point.  It also depicts the problems
 arising when facing the task of calculating the radiation transfer in
 a nonlinear inhomogeneous medium.  We begin with an analysis of a
 system that is not fully in the optically thick limit and continue
 with an estimating the effective opacity when not fully in the thin
 limit.

\subsection{Regions close to the optically thick limit}

 Although it is apparently impossible to solve in a close analytical
 form the opacity reduction in an arbitrary nonlinear configuration,
 we attempt to solve it in at least one realistic configuration in
 order to estimate more accurately $\B$ and $p_h$.  We shall now
 assume for simplicity that the perturbations are vertically
 elongated.  This is actually a reasonable approximation because the
 unstable modes found in Shaviv (\jcite{instabilities}) typically have
 $k_x \sim l_p^{-1}$ but a somewhat smaller $k_z$.  This implies that
 most of the interface between high density and low density regions is
 vertical and that the isothermal contours are mostly horizontal.

 Since the medium is close to the thick limit, we solve for the
 radiation in the vicinity of the interface, assuming that it is flat
 and the two density regions extend infinitely to either side.

 The framed region shown in figure \ref{fig:thick} corresponds to a
 typical area which we wish to solve.  We assume that both sides of
 the region have $B = B_0 + b z$.  The dense region on the right has
 $\rho_h \equiv (1+\delta) \br$ and the underdense on the left has
 $\rho_l \equiv (1-\delta) \br$.

 Using the formal solution given by eq.~(\ref{eq:formalsol}), we can
 write the radiation intensities. For points located in the high
 density region, at a horizontal coordinate $x_0$, the intensity for
 trajectories coming from the positive direction is given by:
\begin{eqnarray}
 I(x_0,\hn_-) \hskip -3pt &=& \hskip -3pt B_0 + b \kappa_0 \br
 {\tan\mu\cos\varphi\over \cos\mu} \\ \nonumber & &\hskip -35pt \times
 \int_{x_0}^{\infty} (1+\delta) (x-x_0) \exp\left[-{\kappa_0 \br
 (x-x_0) (1+\delta) \over \cos\mu}\right] dx,
\end{eqnarray}
 where $\mu$ is the direction of the ray relative to the x-axis
 (perpendicular to the interface) such that $s = x/\cos\mu$ and
 $z=x\tan\mu\cos\varphi$. For trajectories coming from the negative
 x-direction, we have:
\begin{eqnarray}
 I(\vx,\hn_+) \hskip -5pt &=& \hskip -5pt B_0 + b \kappa_0 \br
 {\tan\mu\cos\varphi\over \cos\mu} \\ \nonumber & & \hskip -55pt
 \hfill \times \left\{ \int_{0}^{x_0} (1+\delta) (x-x_0)
 \exp\left(-{\kappa_0 \br (x-x_0) (1+\delta) \over \cos\mu}\right) dx
 \right.  \\& & \hskip -55pt \hfill + \left.\hskip -3pt
 \int_{-\infty}^{0} \hskip -5pt (1-\delta) (x-x_0) \exp \hskip -3pt
 \left[-{\kappa_0 \br \left[x_0(1+\delta)+x(1-\delta) \right]\over
 \cos\mu}\right] \hskip -3pt dx \right\}.
\nonumber
\end{eqnarray}
This time $\cos\mu$ is negative. If we now integrate to get the flux,
we find
\def\tx{{\tilde{x}_0}}
\begin{eqnarray}
 F(x_0>0) \hskip -3pt &=& \hskip -3pt {4\pi b\over 3 c \kappa_0 \br (1
 + \delta) } + { \pi b \over 3 c \br \kappa_0} (1-\delta^2) \\
 \nonumber & & \hfill \hskip -55pt \times \left[ \tx \delta (1+\delta)
 \left( -6 + \tx^2(1+\delta)^2) \Gamma(0,\tx(1+\delta))\right)
 \right. \\ \nonumber && \hskip -55pt \hfill + \left.  \delta (-4 +
 \tx (1 + \delta) (-1 + \tx + \tx \delta)) \exp(-\tx (1+\delta))
 \right]
\end{eqnarray}
 with $\tx\equiv x_0 \kappa_0 \br$. $\Gamma(n,x)$ is the incomplete
 $\Gamma$ function (and should not be confused with the Eddington
 parameter $\Gamma$). The first term is the uncorrected flux $F_h$ in
 the high density region while the rest of the expression is the
 correction near the interface with the low density region. The net
 {\em correction} to the average flux of the system from only
 corrections in the positive x-axis is:
\begin{equation}
 \Delta F_+ = {S\over V_{\mr{tot}}} \int_0^{\infty} (F-F_{h}) dx = {
 \pi (S/V_{\mr{tot}})\delta^2 \over 2 c \kappa_0 \br (1-\delta)
 (1+\delta)^2}.
\end{equation}
 The flux in the negative half, where the density is $\br (1-\delta)$
 is obtained by the transformation $\delta\rightarrow -\delta$. The
 sum of the net correction to the flux on both sides of $x=0$ is the
 total correction to the average flux in the system:
\begin{equation}
 \Delta F = \Delta F_+ + \Delta F_- = - {\pi (S/V_{\mr{tot}}) b
 \delta^2 \over c \kappa_0 \br (1-\delta^2)^2}.
\end{equation}
 The average flux of the system, had it been homogeneous, is $F_0 = 4
 \pi b/3 c \kappa_0 \rho_0$. If we then assume $\Gamma \gg 1$ or
 $\delta \rightarrow 1$ and use eq.~(\ref{eq:F0}), we find:
\begin{equation}
 {\Delta F \over F_0} = \left( \kappa_0 \over
 \kappa_{\mr{eff},\infty}\right)^2 { 3 S \over 32 \alpha (1-\alpha)^2
 \kappa_0 \rho_0 V_{\mr{tot}}}.
\end{equation}

 This result should be compared with the one obtained in the more
 heuristic treatment (eq.~\ref{eq:Fthick}).  The correction to the
 flux obtained here has an additional factor of $3/(8 \alpha)$ but is
 otherwise the same.  That is, if we compare $\kappa_{\mr{eff}}$ as
 given by eq.~(\ref{eq:kappabehavior}) with $\psi=\B
 \Delta\tau^{-p_h}$, we obtain $p_{h}=1$ and $\B=(3 \Xi / 32 \alpha
 (1-\alpha)^2)/ \beta$.

\subsection{Regions close to the optically thin limit}

 We begin with systems approaching the thin limit.  To calculate the
 radiation transfer we should perform a linear analysis in which the
 zeroth order terms are those of the thin limit.  That is to say, the
 zeroth order terms are those arising when the optical depth of the
 perturbations can be neglected (even if $\delta \rho/\rho \sim {\cal
 O}(1)$).

 The formal solution to the radiation transfer is given by
 eq.~(\ref{eq:formalsol}). It assumes that the boundaries are many
 optical depths away. This is valid also in the optically thin case
 because even if a scale height is optically thin, the generated wind
 is optically thick unless $\Gamma-1$ is extremely smaller than unity.
 We also implicitly assume that the photon crossing time is
 significantly shorter than any dynamical time scale in the system on
 which the perturbations can change their structure.

 We write the radiation source function of the unperturbed region as
 $B_{0} (\bf{r})$ and the density as $\rho_{0}(\bf{r})$, while the
 corresponding perturbations are $B_{1}(\bf{r})$ and $
 {\rho_{1}(\bf{r})}$.
 
 If we perturb the formal solution of the transfer equation
 (eq.~\ref{eq:formalsol}) we find:

\begin{eqnarray}
 I_0(\vx,\hn) & = & \int_0^\infty \chi_0 B_0 e^{-\chi_0 s} ds
 \\ \nonumber 
 I_1(\vx,\hn) & \approx& \int_0^\infty \chi_0 B_1 e^{-\chi_0 s}ds +
 \int_0^\infty \chi_1 B_0 e^{-\chi_0 s}ds \\ \nonumber & & -
 \int_0^\infty \chi_0 B_0 e^{-\chi_0 s} \left( \int_0^s \chi_1
 ds'\right) ds.
\end{eqnarray}

 The small parameter used for the expansion is $E_{1} \equiv
 \int_{0}^{s}\chi_{1}(s') ds'$, namely, the perturbation to the
 extinction $E$ which appears in the exponent.  If $E_{1}$ is much
 smaller than unity, than $\exp(-E_{0}-E_{1}) \approx
 \exp(-E_{0})(1-E_{1})$.  Note that $E_{1} \ll 1$ for $s$'s of order
 the m.f.p of a photon or smaller.

\def\BZ{\bar{B}_{0}}
 Next, we assume for simplicity that the unperturbed radiation source
 function can be written as: $B_0({\vx})=\BZ+ \vx\cdot\vb$. After the
 linearization, it is meaningful to look for harmonic solutions to the
 equations of the form $X_1 = {\tilde X}_1 e^{i \vk \cdot \vx}$ where
 $X_{1}$ is any perturbed quantity.
 
 The expression for the first order perturbation to the specific
 intensity can then be obtained rather straightforwardly if we change
 the integration order for the double integral and choose an
 integration coordinate system in which $\vb$ aligns with the $z$-axis
 (as opposed to aligning $\vk$ with $z$). One then finds:

\begin{equation}
I_1(\vx,\hn) = \left( {\chi_0 B_{1} \over i \vk \cdot \hn + \chi_0}
+  {1\over\chi_0} {(\hn\cdot\vb) \chi_1 \over i \vk \cdot \hn +
\chi_0} \right) e^{i \vk \cdot \vx}.
\label{first-order-I}
\end{equation}
 The first term describes changes in the
 source function due to local changes in the temperature.  The second
 term describes changes in the optical depth due to temperature and
 density perturbations along the line of sight.

 Once the expression for the perturbed radiative flux is known, the
 moments of $I$ can be evaluated directly from the perturbed radiation
 field.  Starting with the lowest moment, $J$ is:
\begin{eqnarray}
 J_1(\vx) &\nsk=\nsk& {1\over 4\pi} \oint I_1(\vx,\hn)d\vO \\
 \nonumber &\nsk=\nsk& {1\over 4\pi} \left[ \oint {\chi_0 B_1 d\vO
 \over i \vk \cdot \hn + \chi_0} + {1 \over \chi_0} \oint
 {(\hn\cdot\vb)\chi_1 d\vO \over i \vk \cdot \hn + \chi_0} \right]
 e^{i \vk \cdot \vx} \\ \nonumber &\nsk=\nsk& \left[
 \Lambda_1\left(k\over\chi_0\right) B_1 -i {\vk\cdot\vb\over 3
 \chi_0^2} \Lambda_2\left(k\over\chi_0 \right) {\chi_1\over\chi_0}
 \right] e^{i \vk\cdot\vx}.
\label{eq:jone}
\end{eqnarray}
To obtain the last equality we choose the integration coordinate system
so that $\vk$ is parallel to its $z$-axis. In this particular choice   the 
denominators of the integrals obtain their simplest  form. We
also defined the following three functions for convenience:
$$
\begin{array}[c]{rclrcl}
 \Lambda_1(\xi) & \nsk\equiv\nsk & {1\over\xi}\tan^{-1}(\xi) &
 \Lambda_1(\xi \gg 1) & \nsk\approx\nsk & {\pi\over2} {1\over \xi} 
\\
 \Lambda_2(\xi) & \nsk\equiv\nsk &
 \left({3\over\xi^2}-{3\over\xi^3}\tan^{-1}(\xi)\right) &
 \Lambda_2(\xi \gg 1)& \nsk\approx\nsk &{3\over \xi^2} 
\\
 \Lambda_3(\xi) & \nsk\equiv\nsk &
 {3\left[(\xi^2+1)\tan^{-1}(\xi)-\xi\right] \over 2 \xi^3} &
 \hskip 6pt \Lambda_3(\xi \gg 1) & \nsk\approx\nsk &{3\pi\over4} {1\over 
 \xi}.
\end{array}
$$
 The third function will be used shortly.  The unperturbed value of
 $J$ is $J_{0}=\BZ$.

 The first moment of $I$ is related to the flux (${\bf F}=(4 \pi/c)
 \vH$) and given by:

\begin{eqnarray}
 \vH_1(\vx) &\nsk=\nsk& {1\over 4\pi} \oint I_1(\vx,\hn) \hn d\vO \\
 \nonumber &\nsk=\nsk& {1\over 4\pi}\left[ \oint {\chi_0 \hn B_1 d\vO
 \over i \vk \cdot \hn + \chi_0} + \oint {(\hn\cdot\vb)\hn \chi_1 d\vO
 \over i \vk \cdot \hn + \chi_0} \right] e^{i \vk \cdot \vx} \\
 \nonumber & \nsk=\nsk & \left\{{\vk \over 3 i \chi_0}
 \Lambda_2\left(k\over\chi_0\right) B_1 + {\chi_1\over 3\chi_0^2}
 (\vb\cdot\hk) \hk \Lambda_2\left(k\over\chi_0\right) \nsk\right.  \\
 \nonumber & \nsk & + \left.  {\chi_1\over 3 \chi_0^2}
 (\vb - (\vb\cdot\hk) \hk ) \Lambda_3\left(k\over\chi_0\right)
 \right\} e^{i \vk\cdot\vx}
\label{H1eq}
\end{eqnarray}
where $\hk$ is a unit vector in the direction 
of $\vk$. The first term arises from perturbations to the temperature 
while the second and third from perturbations to the opacity.

 When close to the Eddington limit, the region where the optical depth
 of a scale height is of order unity should be in the NAR limit (e.g.,
 \cite{instabilities}, \cite{Glatzel93}) unless the absorption opacity
 is extremely small (smaller than about
 $(p_{\mr{rad}}/p_{\mr{gas}})(v_s/c)$ times the ratio between the
 absorption opacity and the scattering opacity).  Consequently, one
 can assume that heating or cooling of the gas is instantaneous and
 that $B = J$.  If we impose $J_{1}=B_{1}$ (since the zeroth order
 terms satisfy it anyway), we find from eq.~(\ref{eq:jone}) that:
\begin{equation}
 B_{1} = - {\Lambda_{2}\left(k\over\chi_0\right) \over
 1-\Lambda_{1}\left(k\over\chi_0\right)} {i\vk\cdot \vb \over 3
 \chi_{0}^{2}} {\chi_{1}\over \chi_{0}}.
\end{equation}
If we write $\xi \equiv \left(k/\chi_0\right)$ then the correction 
to the flux is:
\begin{eqnarray}
 \vH_1(\vx) &\nsk=\nsk& \left\{ {\vk \over 3 \chi_0}
 {\Lambda_2^{2}(\xi)\over \Lambda_{1}(\xi)-1}{\vk\cdot \vb \over 3
 \chi_{0}^{2}} + {\chi_1\over 3\chi_0^2} (\vb\cdot\hk) \hk
 \Lambda_2\left(\xi\right) \nsk\right.  \nonumber \\ & \nsk & + \left.
 {\chi_1\over 3\chi_0^2} (\vb - (\hb\cdot\hk) \hk )
 \Lambda_3\left(\xi\right) \right\} e^{i \vk\cdot\vx}.
\label{H1eq2}
\end{eqnarray}
 First, we note that the first two terms are proportional to
 $\xi^{-2}$ for large $\xi$'s or short wavelengths while the third
 term falls only as $\xi^{-1}$ such that it would dominate for low
 $\xi$'s.  This fact provides the justification for neglecting the
 temperature perturbations in the heuristic picture given in
 \S\ref{sec:criticalpoint}.  Moreover, careful analysis of the
 definition of the $\Lambda_{i}$'s will reveal that the first two
 terms actually cancel each other.

 The effective opacity of the medium can now be calculated using
 Shaviv (\jcite{porous}):
\begin{eqnarray}
 \chi_{\mathrm{eff}} &=& {\left< H \chi \right> \over \left< H
 \right>} = {H_{0}\chi_{0} +\left< H_{1}\chi_{1}\right> + \left<
 H_{1}\right>\chi_{0} +H_{0}\left< \chi_{1}\right> \over H_{0} +
 \left< H_{1}\right>} \nonumber \\ &=& {H_{0} \chi_{0}+ \left<
 H_{1}\chi_{1}\right> \over H_{0}} \nonumber \\&=& {\chi_{0}} - {3
 \pi\over 8}{\chi_{1}^{2}\over\chi_{0}}( \hz - (\hz\cdot\hk)
 \hk ) {\chi_{0}\over k}
\end{eqnarray}
 where we have used the $\xi \gg 1$ expansion of $\Lambda_{3}$ and
 assumed without loss of generality that $\vb$ is in the $\hz$
 direction. We point out that had we worked with the opacity per unit
 mass $\kappa$ instead of the extinction $\chi$, which is the opacity
 per unit volume, the averaging $\left<\right>$ would have been a mass
 weighted average and not a volume average.

 It would be interesting to solve the problem exactly in the same case
 which is solved for in the thick limit, namely, that of vertical
 slabs of width $d$ and a density of $\rho_{0}(1\pm\delta)$. To do so,
 we use the Fourier series expansion for the density:
\begin{equation}
    \rho(x)=\rho_{0}\left(1+{ 4 \delta\over \pi}\sum_{n=0}^{\infty}
    {\sin \left[ (2n+1)\pi x / d\right] \over 2 n +1}\right)
\end{equation}
For Thomson atmospheres we find that $\chi_{1}=\kappa_{0}\rho_{1}$.
If the identity $\sum_{n=0}^{\infty}1/(2n+1)^{3}=7 \zeta(3) /8$ is
used (with $\zeta(x)$ being the $\zeta$-function), we obtain:
\begin{equation}
 {\kappa_{\mathrm{eff}}\over\kappa_{0}}={\chi_{\mathrm{eff}}\over
 \chi_{0}} = 1-{21 \zeta(3) d\kappa_{0}\rho_{0}\over 2 \pi^2}
 \delta^{2} \approx 1.28 d \kappa_0 \rho_0 \delta^2.
\end{equation}
 We can also write $\delta^{2} = (\kappa_{0} -
 \kappa_{\mathrm{eff},\infty})/\kappa_{0}$ in the case of vertical
 slabs.  If we now compare this result to eq.~(\ref{eq:kappabehavior}),
 with $\psi(\Delta\tau\ll1) \approx 1-\Delta\tau^{p_{l}}$, we obtain:
 $p_{l}=1$ and $\A\approx 1.28 d \kappa_{0} \rho_{0}/ \Delta \tau =
 1.28 \beta $.
 
 In the more general case of $\alpha \neq 1/2$, an analytic solution
 does exist but it involves poly-logarithm functions of $\alpha$ 
 which would appear in $\A$, changing it by ${\cal O}(1)$.

\label{lastpage}
\end{document}